\documentclass{aastex631}

\usepackage{soul}
\usepackage{longtable}
\usepackage{tabularx}
\usepackage{longfigure}
\usepackage{rotating}
\usepackage[toc,page]{appendix}
\submitjournal{PSJ}

\graphicspath{{./}{figures/}}

\begin{document}

\title{Spectral Effects of Shock Darkening on Ordinary Chondrite and Howardite, Eucrite, and Diogenite meteorites}

\correspondingauthor{Juan A. Sanchez}
\email{jsanchez@psi.edu}

\author{Juan A. Sanchez}
\affiliation{Planetary Science Institute, 1700 East Fort Lowell Road, Tucson, AZ 85719, USA}

\author{Vishnu Reddy}
\affiliation{Lunar and Planetary Laboratory, University of Arizona, 1629 E University Blvd, Tucson, AZ 85721-0092}

\author{Lucille Le Corre}
\affiliation{Planetary Science Institute, 1700 East Fort Lowell Road, Tucson, AZ 85719, USA}

\author{Neil Pearson}
\affiliation{Planetary Science Institute, 1700 East Fort Lowell Road, Tucson, AZ 85719, USA}

\author{Adam Battle}
\affiliation{Lunar and Planetary Laboratory, University of Arizona, 1629 E University Blvd, Tucson, AZ 85721-0092}

\begin{abstract}

Impacts are the most ubiquitous processes on planetary bodies in our solar system. During these impact events, shock waves can deposit enough energy to produce shock-induced 
darkening in the target material, resulting in an alteration of its spectral properties. This spectral alteration can lead to an ambiguous taxonomic classification of asteroids and an 
incorrect identification of meteorite analogs. In this study, we investigated the effects of shock darkening on the visible and near-infrared spectra of ordinary chondrites and 
members of the howardites, eucrites, and diogenites clan. A decrease in albedo and suppression of the absorption bands with increasing shock darkening was observed for 
all the samples. We found that adding $\gtrsim$50\% of shock-darkened material to an unaltered sample was enough to change the taxonomic classification of an ordinary chondrite 
from S-complex to C- or X-complex. A similar amount was sufficient to change the taxonomic classification of a eucrite from V-type to O-type, whereas a eucrite composed of 100\% 
shock-darkened material was classified as a Q-type. We discuss the limitations of using just albedo for taxonomic classification of asteroids, which has implications for future 
space-based infrared surveys. We also investigated if shock darkening is responsible for a bias against the discovery of near-Earth objects (NEOs) with H and L chondrite-like 
compositions, which could explain the abundance of LL chondrites among NEOs.

\end{abstract}


\section{Introduction} \label{sec:intro}

The spectral properties of asteroids can be affected by different processes acting on the surface of these bodies. One such process is the shock-induced darkening 
that can result from meteorite impacts and asteroid collisions. During these impact events, shock waves deposit some of their energy as heat within the target rocks, 
 leading to a rapid increase in pressure and temperature \citep{1998trca.book.....F}. At pressures of $\sim$40-50 GPa and temperatures of $\sim$1463 K shock darkening is produced when troilite, one of 
the opaque minerals present in the rock, melts, forming a network of fine, thin veins within the silicate grains \citep[e.g.,][]{1967Icar....6..189H, 1985RvGeo..23..277R, 1994GeCoA..58.3905B, 2005M&PS...40.1329H, 2017M&PS...52.2375M, 2020A&A...639A.146K}.  At higher 
pressures of $\sim$90-150 GPa and post-shock temperatures over $\sim$1800 K, a second-stage shock darkening occurs, where shock waves deposit sufficient thermal energy to 
produce bulk melting, forming blebs of FeNi metal and troilite that are finely dispersed in the silicates. The material produced during this second stage is referred to as impact 
melt \citep[e.g.,][]{1998trca.book.....F, 2017M&PS...52.2375M, 2020A&A...639A.146K}. Shock darkening and impact melt have been observed in different types of meteorites, predominantly ordinary chondrites \citep[e.g.,][]{1967Icar....6..189H, 1976JGR....81..905G, 1994GeCoA..58.3905B, 2014Icar..237..116R, 2014Icar..228...78K}, but also among howardites, eucrites and diogenites 
(HEDs) \citep[e.g.,][]{2013M&PS...48..771C, 2017GeCoA.204..159L, 2020M&PS...55..781R}.

The finely dispersed molten opaque minerals that are responsible for the darkening of the rock will also produce changes in the visible and 
near-infrared (VNIR) spectra (0.35-2.5 $\mu$m). This spectral alteration is characterized by a decrease in reflectance and suppression of the absorption bands \citep[e.g.,][]{1976JGR....81..905G, 1989LPSC...19..537B, 1991Metic..26..279B, 1994GeCoA..58.3905B, 2014Icar..237..116R}. The spectral effects of impact melt are essentially the same as those produced during the first stage of shock darkening. Thus, for simplicity, in the present work the term "shock darkening" will be used to refer to both processes, and the term "impact melt" will only be 
used to discuss previous work where it was explicitly mentioned.

Despite all of the work done in the past to understand how shock darkening is produced and its effect on the spectral properties of meteorites, it was not until recent 
years that evidence of shock darkening in asteroids started to emerge. The possible presence of shock darkening and impact melt has been found in the main belt and among the near-Earth objects (NEOs) population. Observations of 4 Vesta carried out by the spacecraft Dawn revealed the presence of orange material deposits that could be linked with impact melt \citep{2013Icar..226.1568L}. These deposits included diffuse ejecta in impact craters (such as Octavia and Oppia), lobate patches with well-defined edges, and ejecta rays 
from fresh-looking impact craters \citep{2013Icar..226.1568L}. Octavia and Oppia crater ejecta have been considered a possible source for the dark clasts in the Allan Hills 76005 
polymict eucrite group \citep{2020M&PS...55..781R}. 

\cite{2014Icar..237..116R} observed the Baptistina family and noticed that the NIR spectra of many of the asteroids exhibit subdued absorption bands. The relatively low 
albedo of this family along with the weak absorption bands of the asteroids were attributed to the presence of a significant shock-darkened component in the surface regolith 
of these objects. \cite{2014Icar..237..116R} also demonstrated that adding shock-darkened material to an ordinary chondrite can change its taxonomic type from S-complex 
to C/X-complex. 

\cite{2022PSJ.....3..226B} found evidence of shock darkening in the NEO (52768) 1998 OR2 that 
could explain its weak absorption bands and its classification as an Xn-type. They showed that the VNIR spectrum of 1998 OR2, which resembles an S-type, can be modeled with a 
mixture of $\sim$63\% of shock-darkened material and $\sim$37\% of light-colored material from the H5 chondrite Chergach. The spectrum of 1998 OR2 was also found to be very 
similar to that of the highly shocked ordinary chondrite McKinney. 

\cite{2024PSJ.....5..131S} identified a group of small NEOs (classified as Sx-types) 
with spectral characteristics and compositions consistent with ordinary chondrites, but whose weak absorption bands could lead to an ambiguous classification in the C- or 
X-complex. They determined that shock darkening could be responsible for the attenuation of the absorption bands in some of these objects.

As more asteroids are discovered and studied, further evidence of shock darkening will probably be found. Hence, it becomes important to 
understand how shock-induced optical alterations can affect our interpretation of the telescopic data. In the present study, we investigate the effects of shock darkening on the 
VNIR spectra of a suite of meteorite samples. In particular, we focus on understanding how spectral darkening can affect the albedo, taxonomic classification, and band parameters used 
for compositional analysis. Our study includes the three subgroups of ordinary chondrites (H, L, LL) plus meteorites belonging to the HED clan. These two meteorite classes 
represent $\sim$90\% of all meteorites that fall on Earth.

\section{The sample} \label{sec:sample}

The sample used in this study is composed of ordinary chondrites, eucrites, and diogenites. Ordinary chondrites are the most common type of meteorites that fall on Earth, representing 
about 86\% of all meteorites. Their composition is dominated by olivine, pyroxene, and plagioclase feldspar, also containing smaller amounts of metal and sulfides 
\citep{2002aste.book..653B, 2006mess.book...19W}. They are divided into three subgroups (H, L, LL) based on the abundance of Fe and the ratio of metallic Fe (Fe$^{0}$) to oxidized Fe (FeO). These meteorites are associated with asteroids in the S-complex \citep[e.g.,][]{1993Icar..106..573G, 2002aste.book..653B, 2011ScienceNakamura, 2019Icar..324...41B, 2024PSJ.....5..131S}. 

Eucrites are basaltic rocks composed mainly of Ca-rich plagioclase feldspar, augite, and low-Ca clinopyroxene (pigeonite). Diogenites are orthopyroxene-rich rocks that formed deeper than the eucrites in 
the lower crust and upper mantle and cooled slowly. Howardites, the other members of the HED clan, are physical mixtures of eucrites and diogenites \citep[e.g.,][]{1998LPI....29.1220M, 2015ChEG...75..155M}. HEDs are linked to V-type asteroids and 4 Vesta based on their spectral similarities and derived composition \citep[e.g.,][]{1970Sci...168.1445M, 1977GeCoA..41.1271C, 1995Icar..115..374H, 2009MandPS...44.1331B, 2010Icar..208..773M}. 

The silicate darkening observed in some ordinary chondrites and HEDs can be attributed not only to shock and heating, but also to a combination of shock and 
regolith processes such as those seen in gas-rich meteorites. Because of this, depending on the darkening mechanism, these meteorites can be grouped into 
different classes. \cite{1991Metic..26..279B} distinguished two types of darkened ordinary chondrites, the 
black chondrites and the gas-rich chondrites. Black chondrites are thought to have formed at the bottoms and sides of impact craters, where they experienced shock and heating during 
the cratering event \citep{1992Icar...98...43K, 1991Metic..26..279B, 1994GeCoA..58.3905B}. These meteorites can have a completely darkened appearance, often displaying small light clasts 
intimately mixed with the darker material. They can also show a light/dark structure where the darkened material is only present in some parts of the meteorite \citep{1991Metic..26..279B, 1994GeCoA..58.3905B}. Gas-rich chondrites, on the other hand, are breccias that typically show light unaltered clasts embedded in a dark matrix. This dark matrix is composed of gas-rich and non-gas-rich grains that are intimately mixed with small clasts \citep[e.g.,][]{1985Metic..20..331W, 1988LPSC...18..573B, 1991Metic..26..279B, 1994GeCoA..58.3905B}. The darkening of 
the matrix is due to a combination of shock and exposure to regolith processes on the surface of the parent bodies \citep{1991Metic..26..279B, 1994GeCoA..58.3905B}. In HEDs, shock darkening has a similar appearance to what we see in ordinary chondrites, i.e., a light/dark structure or light clasts intimately mixed 
with a dark matrix. Like ordinary chondrites, gas-rich HEDs have been also found \citep[e.g.,][]{1967GeCoA..31.1441M, 2011LPI....42.1984W, 2013M&PS...48..771C, 2017GeCoA.204..159L, 2020M&PS...55..781R}.

\begin{figure*}[!ht]
\begin{center}
\includegraphics[height=14cm]{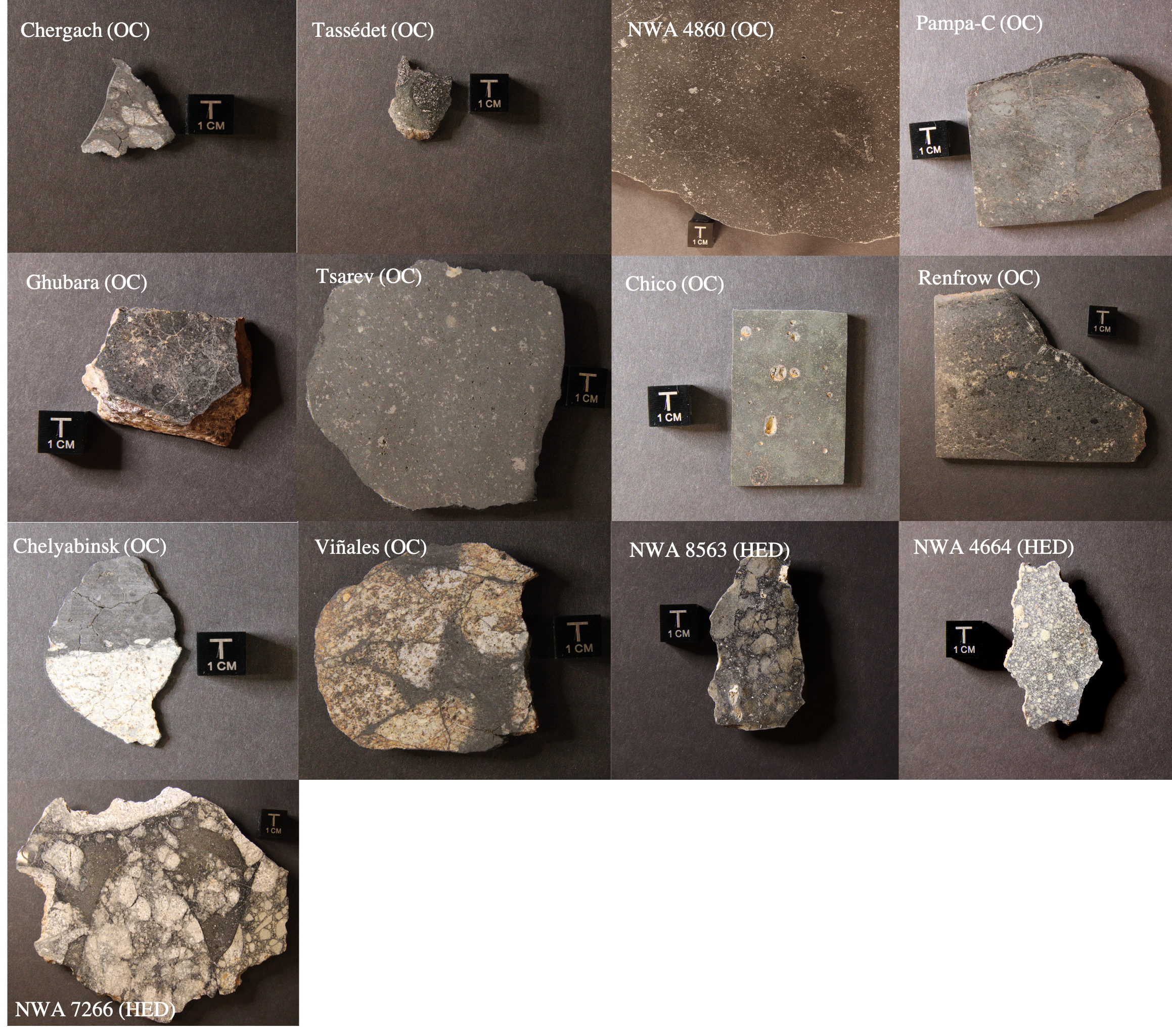}

\caption{\label{f:Samples}{\small Ordinary chondrites (OC) and meteorites belonging to the HED clan used in the present study.}}

\end{center}
\end{figure*}

Our sample includes 10 ordinary chondrites belonging to the three different subclasses, two eucrites, and one diogenite (Figure \ref{f:Samples}). These samples are 
part of our collection curated at the Space Materials Curation Facility at the University of Arizona. Some of these meteorites show signs of terrestrial weathering; however, this does not represent a problem because it is mostly present on the surface. Since there was enough material to work with, it was possible for us to avoid terrestrial weathering products for the most 
part and, if rust was present, it was removed during sample preparation. A description of each sample is presented below.

{\it{Chergach}}. This meteorite fell in El Mokhtar, Erg Chech, Timbuktu district, Mali on July 2007. The meteorite contains H5 chondritic clasts embedded in a fine-grained, black, impact melt matrix with
abundant droplets of metal and metal-troilite \citep{2008M&PS...43.1551W}. The shock stage is S3, and the weathering grade W0. The olivine and pyroxene chemistries, which are given by the mol\% of fayalite (Fa) and ferrosilite (Fs), respectively, are 
Fa$_{18.2}$ and Fs$_{15.5}$ \citep{2008M&PS...43.1551W}. The sample available for this study exhibits light-gray clasts in a dark matrix.

{\it{Tassédet}}. This was found in Nigeria in 2001 and it is classified as an H5 ordinary chondrite \citep{2003M&PS...38..189R}. This melt breccia contains recrystallized melt rock veins and has a shock stage 3 and a weathering grade W1. The olivine and pyroxene chemistries are Fa$_{17.3-17.8}$ and Fs$_{15.6}$, respectively \citep{2003M&PS...38..189R}. The sample used in this study is dominated by dark material.

{\it{NWA 4860}}. This was found in 2003 in northwest Africa. This meteorite is classified as an L4, and it contains heavily shocked chondrules 
surrounded by melt \citep{2009M&PS...44.1355W}. The shock stage is S6, and the weathering grade W1. The olivine and pyroxene chemistries are 
Fa$_{23.7}$ and Fs$_{20.5}$, respectively \citep{2009M&PS...44.1355W}. The sample used in the study only exhibits a shock-darkened phase.

{\it{Pampa-C}}. This was found in the Atacama Desert near Mejillones, Antofagasta, Chile in 1986 \citep{1989Metic..24...57G}. The meteorite was originally classified as an L6 
by \cite{1989Metic..24...57G}, but later reclassified as an L4 with a shock stage 6 \citep{1995Metic..30..785Z}. It is highly weathered and contains numerous fragments of 
chondrules and aggregates \citep{1995Metic..30..785Z}. The olivine present (Fo$_{75.3}$) showed planar deformation features, melt veins, and pockets \citep{1995Metic..30..785Z}. Our sample shows a relatively uniform dark-gray color.

{\it{Ghubara}}. This meteorite was discovered in Oman, Southern Arabia, in 1954. Ghubara is a regolith breccia classified as an L5 chondrite. This black, xenolithic meteorite has a shock stage S4 and a weathering grade W2. The unequilibrated host portion of the olivine grains has a composition of Fa$_{22-27}$, whereas the equilibrated 
xenoliths have a composition of Fa$_{24}$ \citep{1968GeCoA..32..299B}. Ghubara contains both light-colored and dark-colored lithologies \citep{2014M&PS...49..576M}; 
however, the sample available for the present study only contains a dark lithology. 

{\it{Tsarev}}. This was found in Russia in 1968 and possibly fell in 1922. This meteorite is a highly shocked L5 ordinary chondrite \citep{1981Metic..16..193G} 
with olivine and pyroxene chemistries of Fa$_{24.3}$ and Fs$_{20.3}$, respectively \citep{1996M&PS...31..767L}. Tsarev shows both a gray and a black lithology, with the latter 
being the most shocked area of the meteorite, saturated by fine-grained metal and troilite \citep{1989LPSC...19..537B}. The sample analyzed in the present study exhibits mostly a dark matrix with some small and dispersed light clasts.

{\it{Chico}}. This impact melt breccia was found in the Chico Hills region of northeastern New Mexico in 1954 \citep{1954Metic...1..182L}. Chico is a highly shocked (S6) L6 
chondrite consisting of approximately 60 vol\% impact melt and 40 vol\% host chondrite \citep{1995GeCoA..59.1383B}. The meteorite exhibits shock products including melt 
pockets, melt dikes, metallic blebs, and sulfide filled vesicles \citep{1995GeCoA..59.1383B, 1997JGR...10221589Y}. The main minerals present are olivine (Fa$_{25.2}$), orthopyroxene (Fs$_{21.1}$En$_{76.9}$Wo$_{2.0}$), FeNi metal and troilite \citep{1997JGR...10221589Y}. The sample available for this study shows a dull gray color.

{\it{Renfrow}}. This was found in Grant County, Oklahoma, USA, in 1986. This meteorite is classified as an L6 ordinary chondrite; it has a shock grade S5 and an olivine chemistry of 
Fa$_{23.5-24.9}$ \citep{1997M&PS...32..159G}. The sample is dominated by a black matrix. 

{\it{Viñales}}. This fell in Pinar del Rio, Cuba, in 2019. Classified as an L6 chondrite, Viñales has a light-colored interior containing black inclusions of melt rock with silicates 
transected by dark, pseudotachylite-like shock veins \citep{2020M&PS...55.1146G}. The shock stage is S3-S4 and the weathering grade is W0. The main minerals present 
include olivine (Fa$_{24.7}$), low-Ca pyroxene (Fs$_{21.0}$), plagioclase, and troilite. The sample used in the present study exhibits both a light-colored and a dark-colored 
lithology. 

{\it{Chelyabinsk}}. This meteorite fell in Russia in 2013. It is classified as an LL5 ordinary chondrite composed of a light-colored lithology with a typical chondritic texture. 
A dark, fine-grained impact melt lithology is also present \citep{2015M&PS...50.1662R}. The shock stage is S4 and the weathering grade is W0.  The main phases in 
Chelyabinsk are olivine (Fa$_{27.9}$) and orthopyroxene (Fs$_{22.8}$). For the present study, a sample containing both lithologies was used. 

{\it{NWA 7266}}. This meteorite was found in 2012, and it is classified as a monomict eucrite melt breccia. The shock stage and weathering grade are low. NWA 
7266 shows signs of fragmentation and partial melting, and it is composed of rounded clasts of ophitic-textured basaltic eucrite in a deep-brown, fine-grained 
matrix \citep{2015M&PS...50.1662R}. Clasts consist of exsolved pigeonite, calcic plagioclase, silica polymorph, ilmenite, and troilite. \cite{2015M&PS...50.1662R} 
found exsolution textures of thin lamellae of augite (Fs$_{25.3-26.6}$Wo$_{43.8-43.3}$) within the host orthopyroxene (Fs$_{55.8-60.3}$Wo$_{5.6-2.4}$). 
The sample available for this study contains light-colored and dark-colored areas.

{\it{NWA 8563}}. This was found in Mauritania in 2014. NWA 8563 belongs to the HED clan; in particular, it is classified as a monomict eucrite. The shock stage is high and 
the weathering grade is low. The meteorite shows a monomict, subophitic basaltic breccia of eucritic clasts separated by shock veins \citep{2017M&PS...52.1014R}.  Low-Ca 
pyroxene (Fs$_{60.9}$), augite, and plagioclase are the main minerals present \citep{2017M&PS...52.1014R}. The sample available for this study contains light-brown 
clasts embedded in a dark matrix, similar to what is seen in some gas-rich meteorites \citep[e.g.,][]{1988LPSC...18..573B, 1994GeCoA..58.3905B}.

{\it{NWA 4664}}. This was found in Algeria in 2006. This meteorite is a moderately shocked diogenite polymict breccia \citep{2008M&PS...43..571C}. It is composed of 
$\sim$92\% orthopyroxene (Fs$_{17.8-23}$), 4\% olivine (Fa$_{24.1-32.6}$), 2\% plagioclase,  2\% augite, and less than 1\% of chromite and troilite \citep{2008M&PS...43..571C}. The sample used in the present study shows minimal weathering, and it is dominated by very small light clasts intimately mixed 
with dark material.

\section{Methodology} \label{Methodology}

\subsection{Spectral Measurements}

For those samples that contain well-defined and not intimately mixed light-colored and dark-colored areas, both lithologies were carefully separated
under the microscope. These samples include Chergach, Chelyabinsk, Viñales, and NWA 7266. For NWA 8563, we were able to separate a small piece of the light-brown clasts, but it 
was not possible to isolate the dark matrix since it is too intimately mixed with the light-colored material. In the case of NWA 4664, the light clasts present are 
too small and intimately mixed with the dark material, and it was not possible to separate them (see Figure \ref{f:Samples}). All samples were crushed with a pestle and mortar and dry sieved to a grain size of $<$45 $\mu$m. For meteorites Chergach, Chelyabinsk, Viñales, and NWA 7266, three different powders 
were prepared, one with light-colored material, one with dark-colored material, and one intimate mixture of 50 wt\% light-/dark-colored lithologies. For NWA 8563, two different powders were 
prepared, one made from the light-brown clast and one corresponding to the clasts plus dark matrix. For the rest of the meteorites, only one powder was prepared from a representative piece of 
the meteorite. The powders were gently poured into sample cups, and the edge of a glass slide was drawn across the sample to provide a flat surface for the 
spectral measurements.

All data were collected at the Lunar and Planetary Laboratory at the University of Arizona. VNIR spectra (0.35-2.5 $\mu$m) were obtained relative to a Labsphere Spectralon disk using an Analytical Spectral Devices (ASD) LabSpec 4 with a spectral resolution of 3 nm from 0.35-1.0 $\mu$m and 6 nm from 1.0-2.5 $\mu$m. The light source used was a custom quartz tungsten halogen lamp. Spectra were collected at an incident angle $i$ = 0° and emission angle $e$ = 30°. For each measurement, 3000 scans were obtained and averaged to create the final spectrum. 

Ordinary chondrite and HED spectra show absorption bands at $\sim$1 and 2 $\mu$m. The absorption bands in ordinary chondrites are due to the minerals 
olivine and pyroxene, whereas those in HEDs are due to the presence of pyroxene. Olivine has three overlapping bands centered near 1.04-1.1 $\mu$m and the two absorption bands in pyroxene are centered near 0.9-1 $\mu$m and 1.9-2 $\mu$m \citep{1974JGR....79.4829A, 1993macf.book.....B}. These absorption bands are attributed to electronic transitions of Fe$^{2+}$ occupying both the M1 and M2 crystallographic sites in olivine, and the M2 site in pyroxene \citep{1993macf.book.....B}.

\subsection{Band Parameters and Classification}

Spectral band parameters including band centers, band depths, Band Area Ratio (BAR), and spectral slopes were measured using a Python code following the procedure described in \cite{2020AJ....159..146S}. Band centers were measured after dividing out the linear continuum by fitting third- and fourth-order polynomials over the bottom of the absorption bands. 
Band depths were measured from the continuum to the band center and are given as percentage depths. Band areas are defined as the area between the linear continuum and the data curve and 
are calculated using trapezoidal numerical integration. A cutoff at 2.45 $\mu$m was used for the Band II area to minimize the effects of the low signal-to-noise ratio (S/N) at longer 
wavelengths. The BAR was calculated as the ratio of the area of Band II to that of Band I. The spectral slope was calculated as the slope of a linear fit performed 
between the reflectance maxima at $\sim$0.74 and 1.5 $\mu$m. The uncertainties associated with the band parameters are given by the standard deviation of the mean calculated from multiple measurements of each band parameter. Spectral band parameters are shown in Table 1. 

In order to investigate the effects of shock darkening on taxonomic classification we applied the Bus-DeMeo Taxonomy \citep{2009Icar..202..160D} to the meteorite spectra. This 
taxonomic system uses principal components analysis (PCA) to classify asteroids based on their visible and NIR spectra (0.45-2.45 $\mu$m). For this, the Bus-DeMeo 
Taxonomy Classification Web tool$\footnote{http://smass.mit.edu/busdemeoclass.html}$ was used. The taxonomic types assigned to each sample are presented in Table 2.

\begin{deluxetable*}{cccccccccc}

\tablecaption{\label{t:Table1}{\small Spectral band parameters for the meteorite samples. The columns in this table are: name, type, fraction of shock-darkened material (SD), 
albedo (reflectance at 0.55 $\mu$m), spectral slope, Band I center (BIC), Band II center (BIIC), Band I depth (BID), Band II depth (BIID) and Band Area Ratio (BAR).}}

\tablehead{Name&Type&SD ($\%$)&Albedo&Slope ($\mu m^{-1}$)&BIC ($\mu m$)&BIIC ($\mu m$)&BID ($\%$)&BIID ($\%$)&BAR \\ }

\startdata
Chergach&H5&0&0.195&-0.0022$\pm$0.0001&0.921$\pm$0.001&1.925$\pm$0.003&11.2$\pm$0.1&5.0$\pm$0.1&1.01$\pm$0.04 \\
Chergach&H5&50&0.118&-0.0020$\pm$0.0003&0.919$\pm$0.001&1.951$\pm$0.005&7.9$\pm$0.1&4.3$\pm$0.1&1.27$\pm$0.05 \\
Chergach&H5&100&0.075&0.0007$\pm$0.0001&0.917$\pm$0.001&2.096$\pm$0.012&3.6$\pm$0.1&2.9$\pm$0.1&1.52$\pm$0.09 \\
Tassédet&H5&100&0.140&0.0001$\pm$0.0001&0.933$\pm$0.001&1.894$\pm$0.005&7.4$\pm$0.1&3.1$\pm$0.1&1.02$\pm$0.05 \\
NWA 4860&L4&100&0.119&-0.0037$\pm$0.0002&0.934$\pm$0.001&1.980$\pm$0.008&9.2$\pm$0.1&4.0$\pm$0.1&0.82$\pm$0.02 \\
Pampa-C&L4&100&0.080&0.0066$\pm$0.0001&0.938$\pm$0.001&1.992$\pm$0.012&5.5$\pm$0.1&2.6$\pm$0.1&1.11$\pm$0.07 \\
Ghubara&L5&100&0.068&-0.0068$\pm$0.0001&0.942$\pm$0.003&-&5.0$\pm$0.1&-&- \\
Tsarev&L5&100&0.107&-0.0025$\pm$0.0001&0.932$\pm$0.001&2.016$\pm$0.006&7.1$\pm$0.1&3.5$\pm$0.1&0.92$\pm$0.04 \\
Chico&L6&0&0.254&0.0362$\pm$0.0004&0.942$\pm$0.001&1.931$\pm$0.008&21.4$\pm$0.1&8.0$\pm$0.2&0.60$\pm$0.02 \\
Renfrow&L6&100&0.140&0.0232$\pm$0.0001&0.930$\pm$0.001&1.940$\pm$0.006&11.0$\pm$0.2&4.7$\pm$0.1&0.83$\pm$0.05 \\
Viñales&L6&0&0.192&0.0030$\pm$0.0004&0.930$\pm$0.001&1.933$\pm$0.004&18.4$\pm$0.1&6.6$\pm$0.1&0.62$\pm$0.02 \\
Viñales&L6&50&0.122&0.0025$\pm$0.0003&0.927$\pm$0.001&1.957$\pm$0.002&9.6$\pm$0.1&3.5$\pm$0.1&0.72$\pm$0.01 \\
Viñales&L6&100&0.072&0.0040$\pm$0.0002&0.926$\pm$0.001&2.003$\pm$0.014&3.7$\pm$0.1&2.3$\pm$0.1&1.06$\pm$0.08 \\
Chelyabinsk&LL5&0&0.229&0.0060$\pm$0.0005&0.947$\pm$0.001&1.947$\pm$0.007&20.2$\pm$0.1&6.3$\pm$0.1&0.46$\pm$0.01 \\
Chelyabinsk&LL5&50&0.132&0.0034$\pm$0.0003&0.944$\pm$0.001&1.973$\pm$0.005&11.2$\pm$0.1&3.8$\pm$0.1&0.55$\pm$0.02 \\
Chelyabinsk&LL5&100&0.085&0.0024$\pm$0.0001&0.994$\pm$0.002&2.020$\pm$0.006&4.9$\pm$0.1&1.8$\pm$0.1&0.62$\pm$0.03 \\
NWA 7266&Eucrite&0&0.383&0.0722$\pm$0.0006&0.941$\pm$0.001&2.001$\pm$0.001&51.6$\pm$0.1&33.5$\pm$0.1&1.29$\pm$0.01 \\
NWA 7266&Eucrite&50&0.276&0.0420$\pm$0.0004&0.959$\pm$0.001&2.052$\pm$0.001&40.1$\pm$0.1&19.4$\pm$0.1&1.01$\pm$0.02 \\
NWA 7266&Eucrite&100&0.198&0.0302$\pm$0.0001&0.973$\pm$0.001&2.124$\pm$0.002&34.7$\pm$0.1&12.1$\pm$0.2&0.71$\pm$0.03 \\
NWA 8563&Eucrite&0&0.296&0.0545$\pm$0.0007&0.953$\pm$0.001&2.032$\pm$0.003&39.4$\pm$0.1&20.0$\pm$0.1&1.18$\pm$0.01 \\
NWA 8563$^{*}$&Eucrite&-&0.290&0.0547$\pm$0.0006&0.958$\pm$0.001&2.057$\pm$0.003&38.0$\pm$0.1&17.5$\pm$0.1&1.08$\pm$0.01 \\
NWA 4664$^{*}$&Diogenite&-&0.269&0.0523$\pm$0.0004&0.921$\pm$0.001&1.900$\pm$0.002&37.8$\pm$0.1&22.8$\pm$0.1&1.75$\pm$0.02 \\
\enddata
\tablenotetext{*}{These samples contain light clasts intimately mixed with dark material.}
\end{deluxetable*}

\section{Results} \label{sec:Results}

\subsection{Ordinary chondrites}

\begin{figure} 
\includegraphics[width=9.5cm,angle=0]{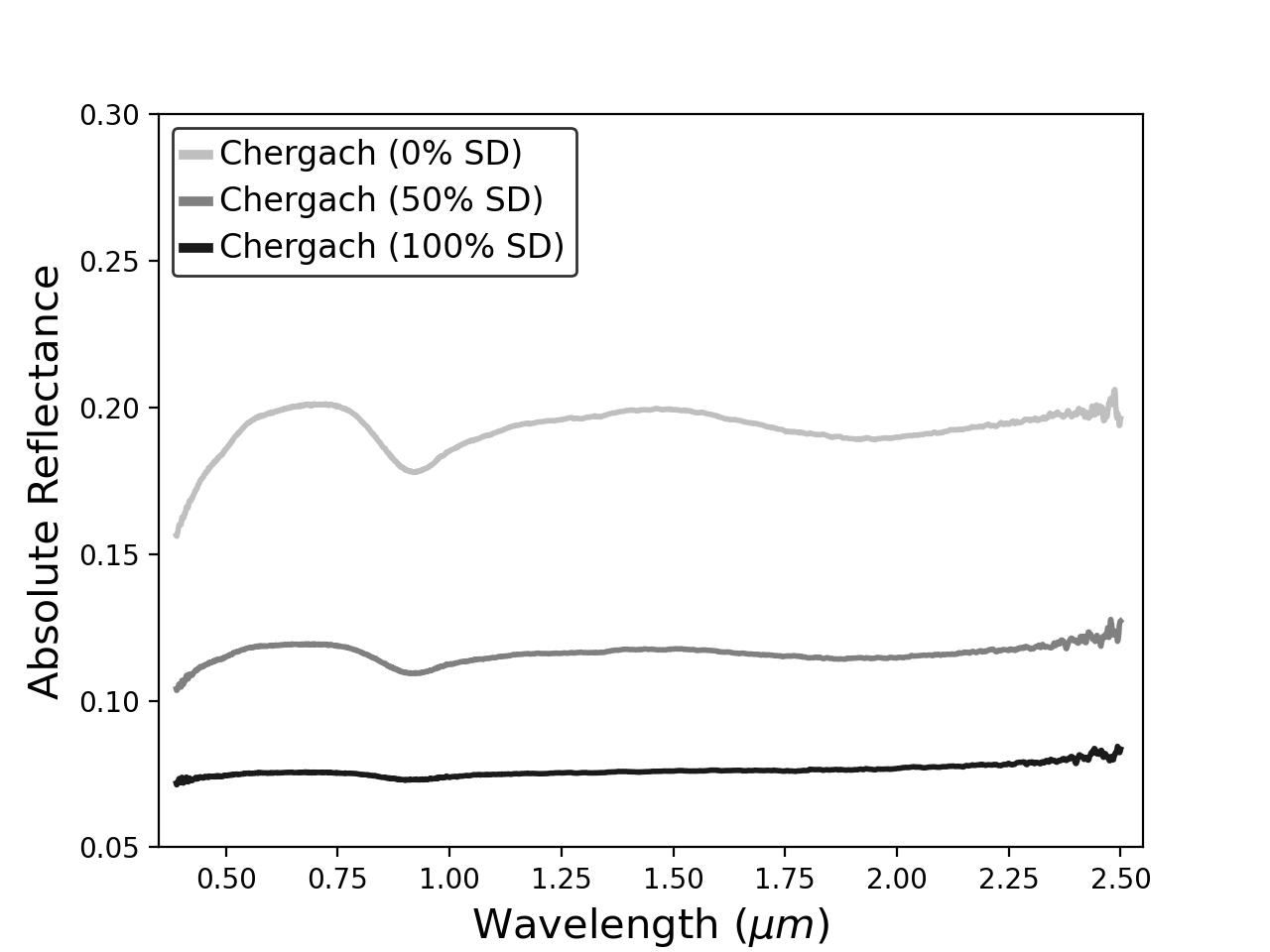}
\hspace{-5mm}
\includegraphics[width=9.5cm,angle=0]{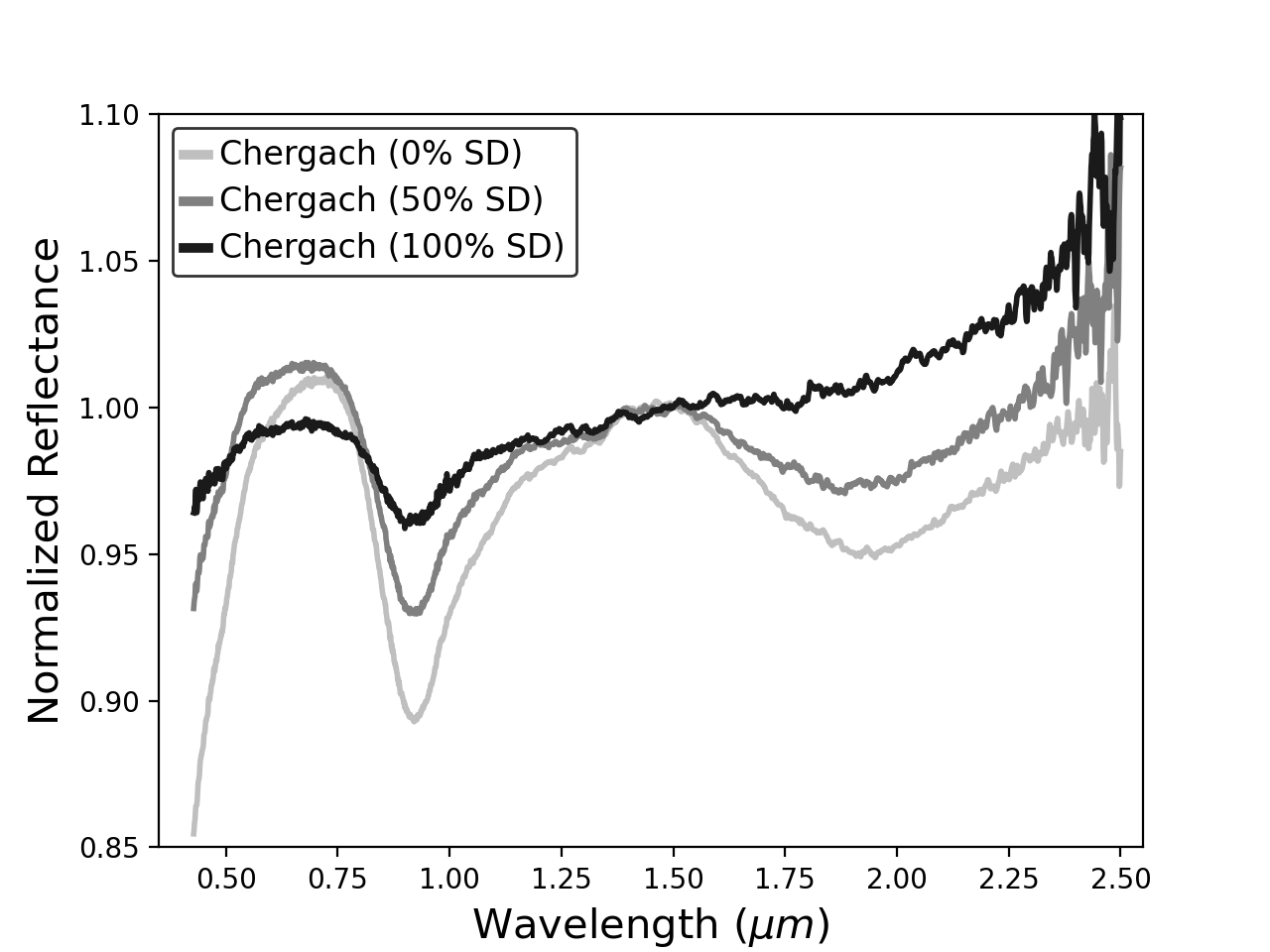}
\caption{Left: Spectra of the H5 ordinary chondrite Chergach corresponding to 0\%, 50\%, and 100\% shock darkening (SD). Right: The same spectra normalized to unity at 1.5 $\mu$m.}

\label{f:Chergach_spec}
\end{figure}

\begin{figure} 
\includegraphics[width=9.5cm,angle=0]{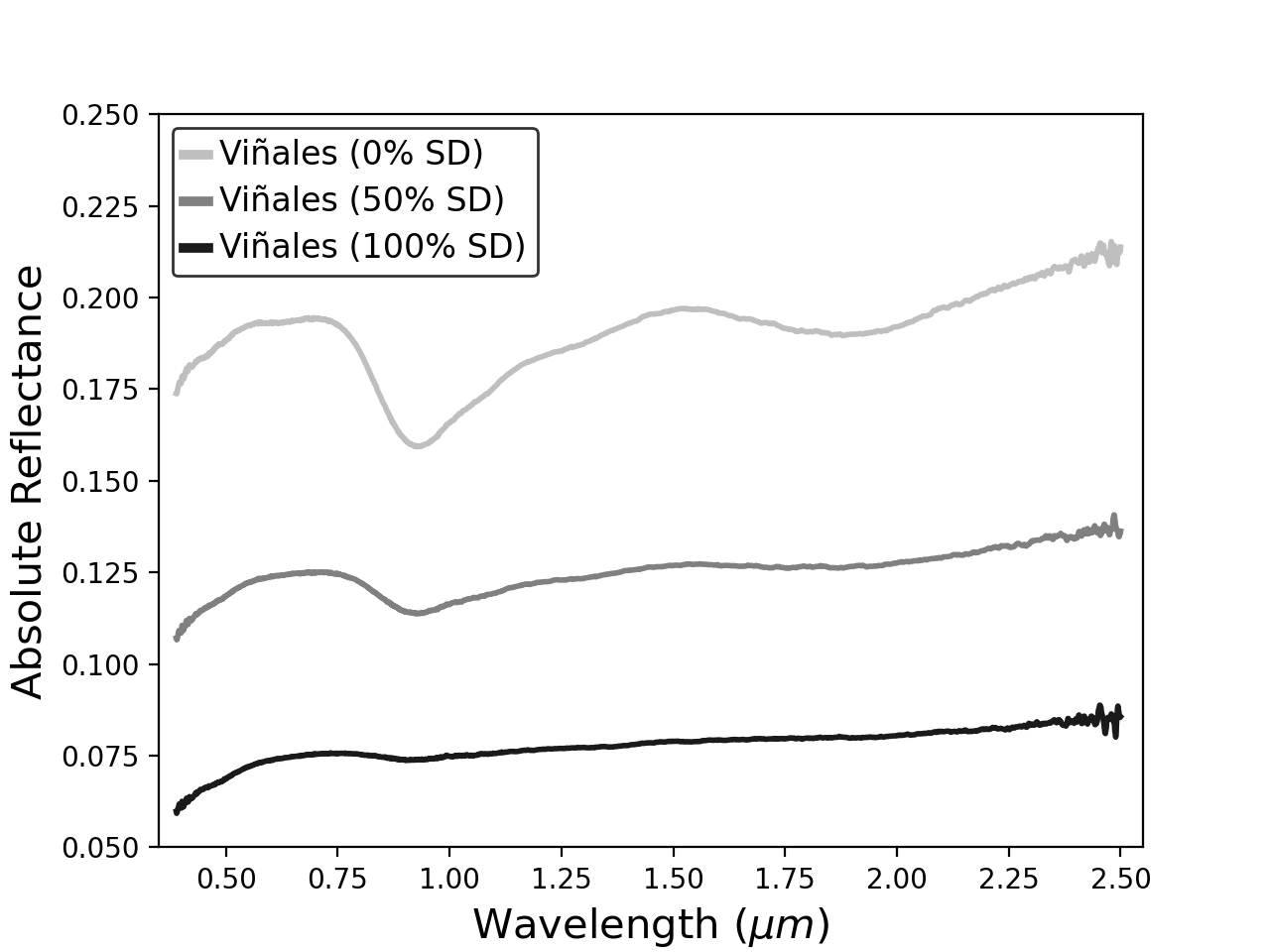}
\hspace{-5mm}
\includegraphics[width=9.5cm,angle=0]{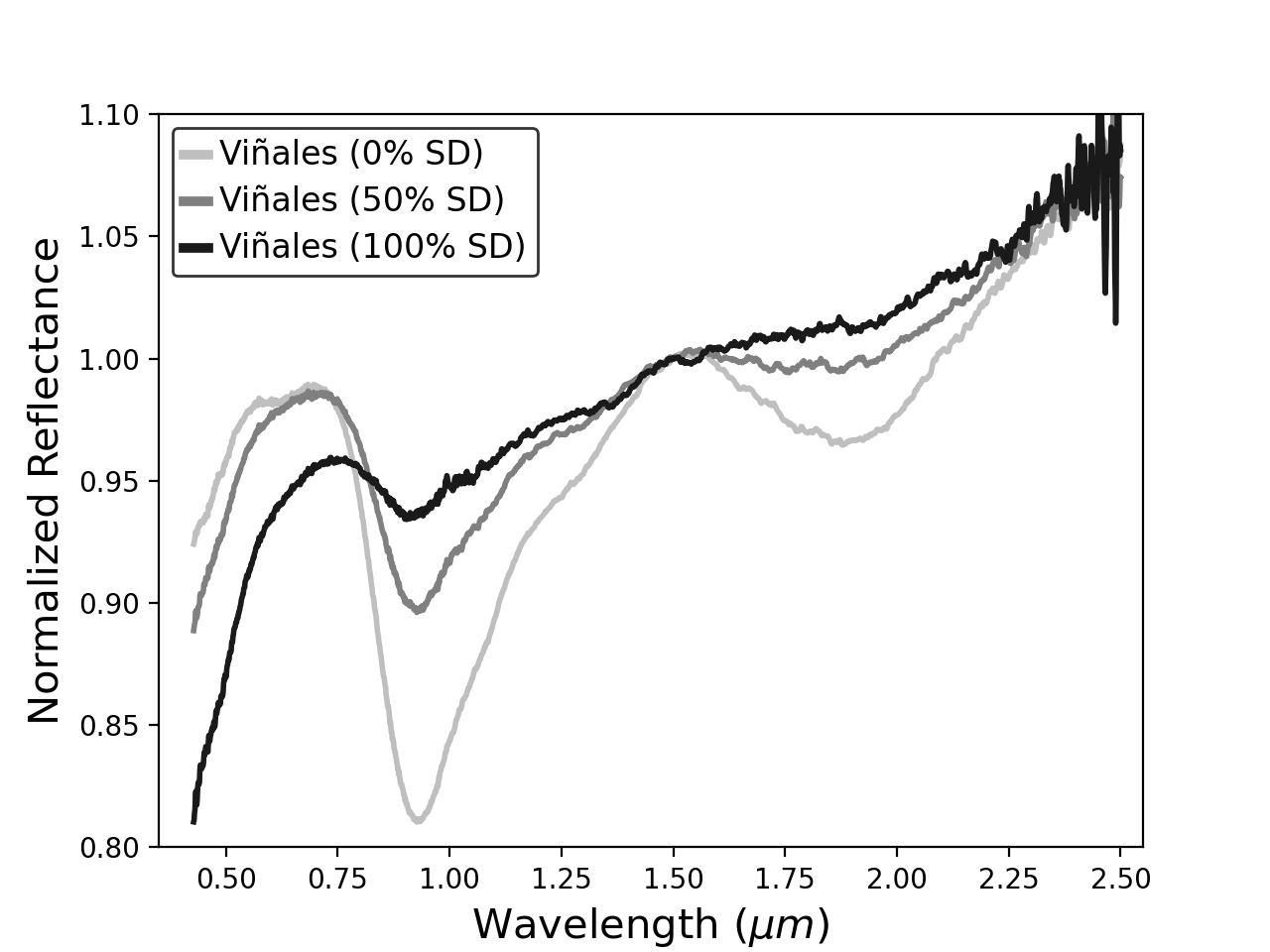}
\caption{Left: Spectra of the L6 ordinary chondrite Viñales corresponding to 0\%, 50\%, and 100\% shock darkening (SD). Right: The same spectra normalized to unity at 1.5 $\mu$m.}

\label{f:Vinales_spec}
\end{figure}

\begin{figure} 
\includegraphics[width=9.5cm,angle=0]{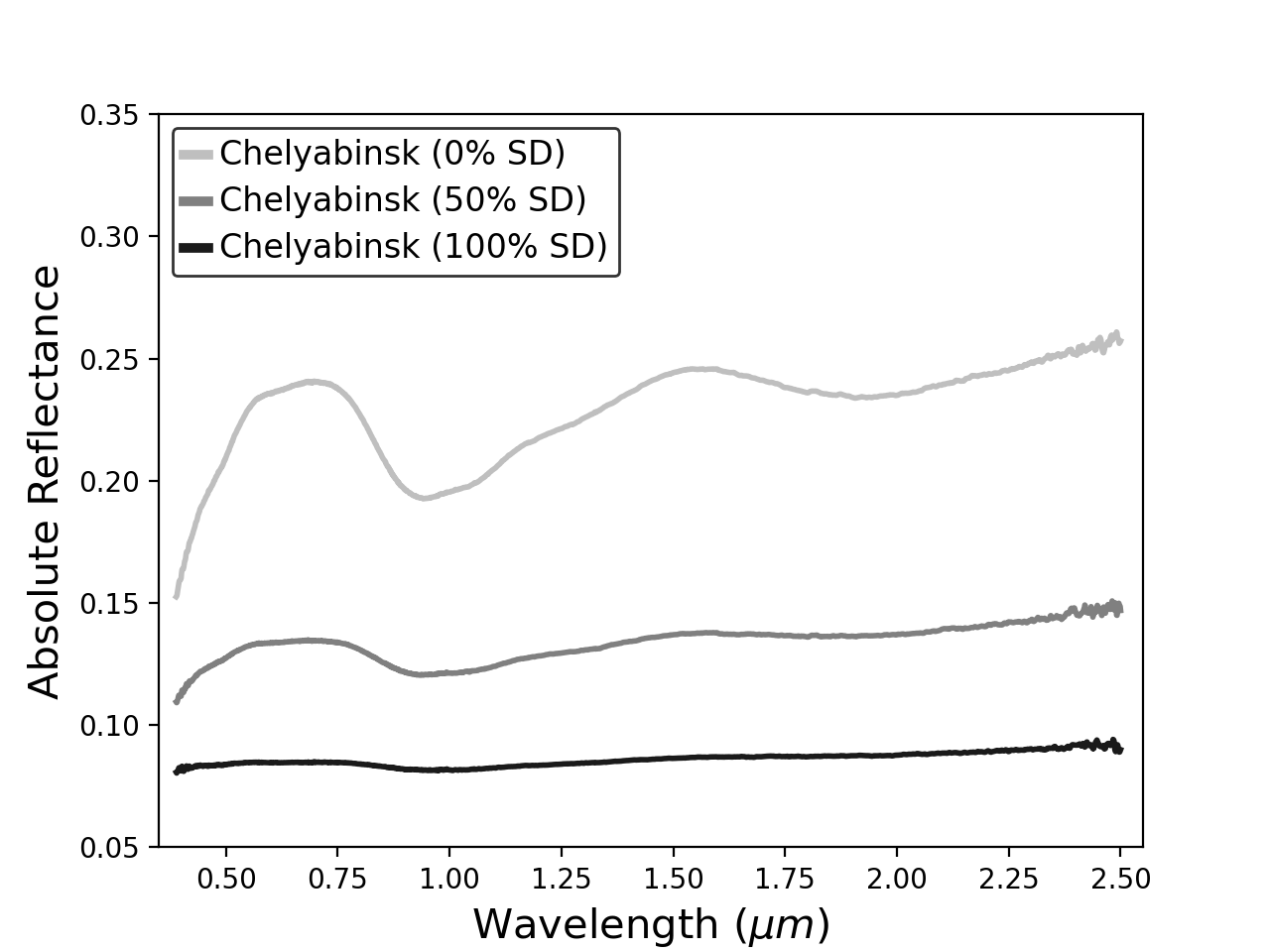}
\hspace{-5mm}
\includegraphics[width=9.5cm,angle=0]{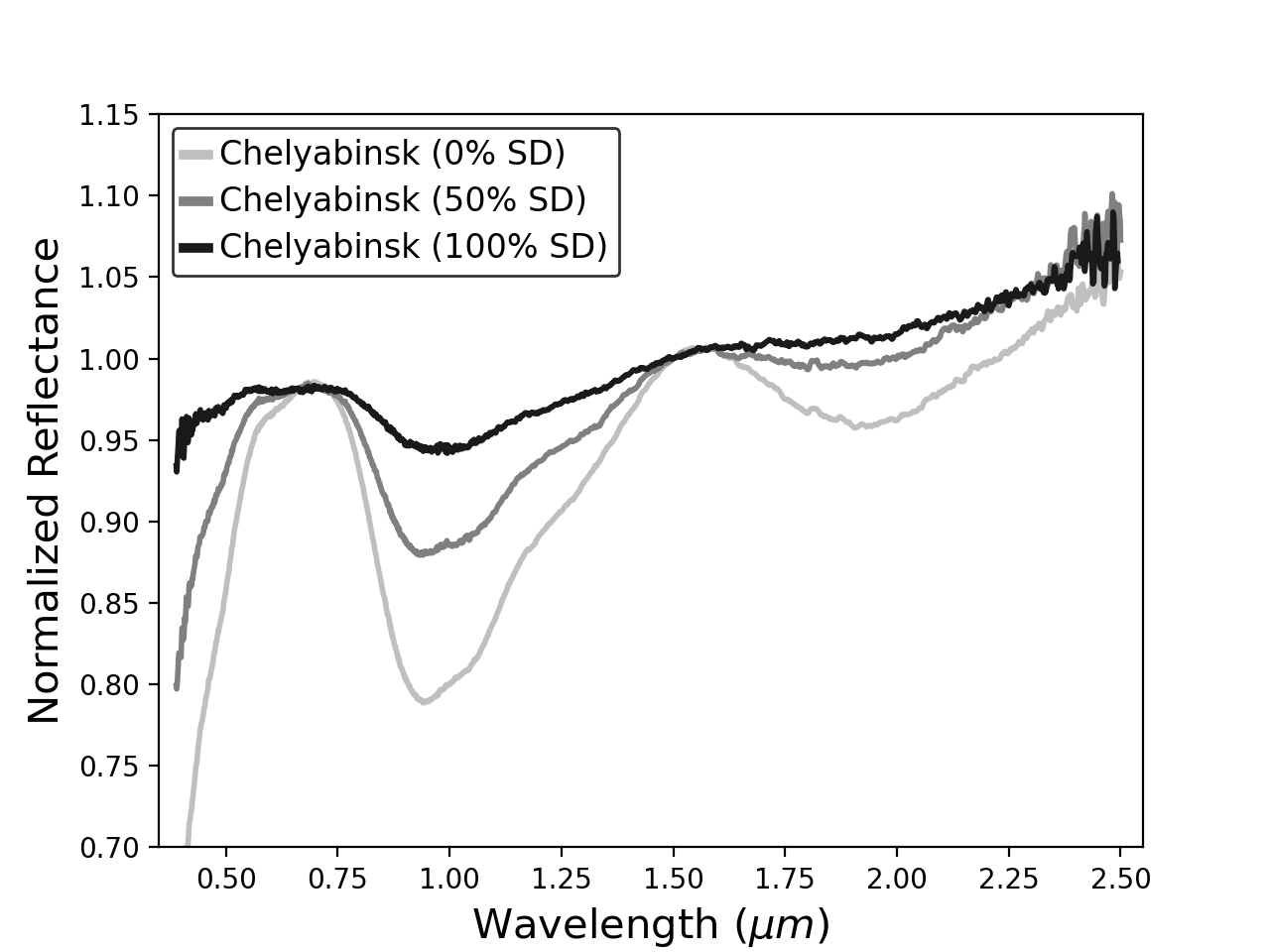}
\caption{Left: Spectra of the LL5 ordinary chondrite Chelyabinsk corresponding to 0\%, 50\%, and 100\% shock darkening (SD). Right: The same spectra normalized to unity at 1.5 $\mu$m.}

\label{f:Chelyabinsk_spec}
\end{figure}

\begin{figure} 
\includegraphics[width=9cm,angle=0]{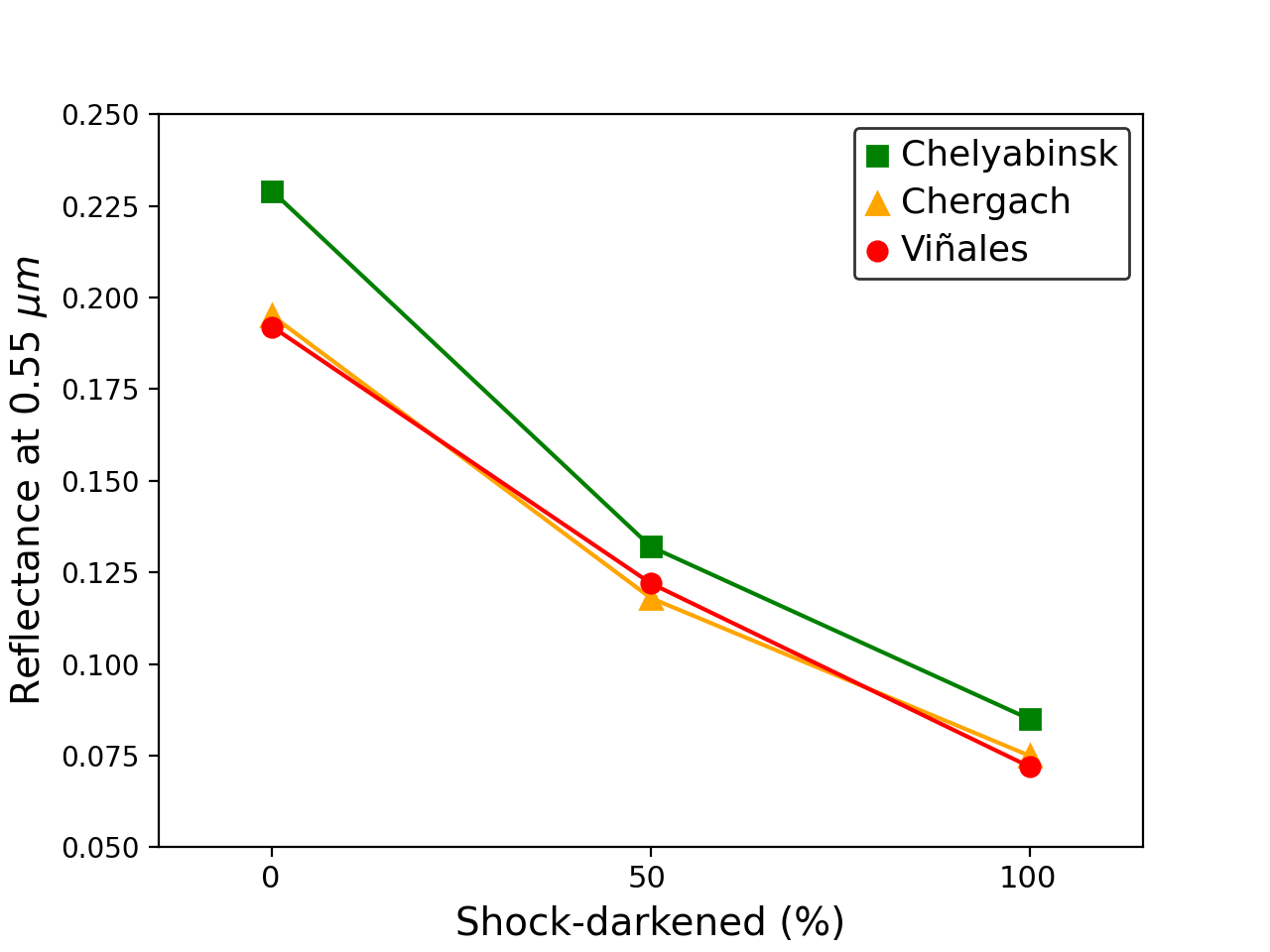}
\hspace{-5.5mm}
\includegraphics[width=9cm,angle=0]{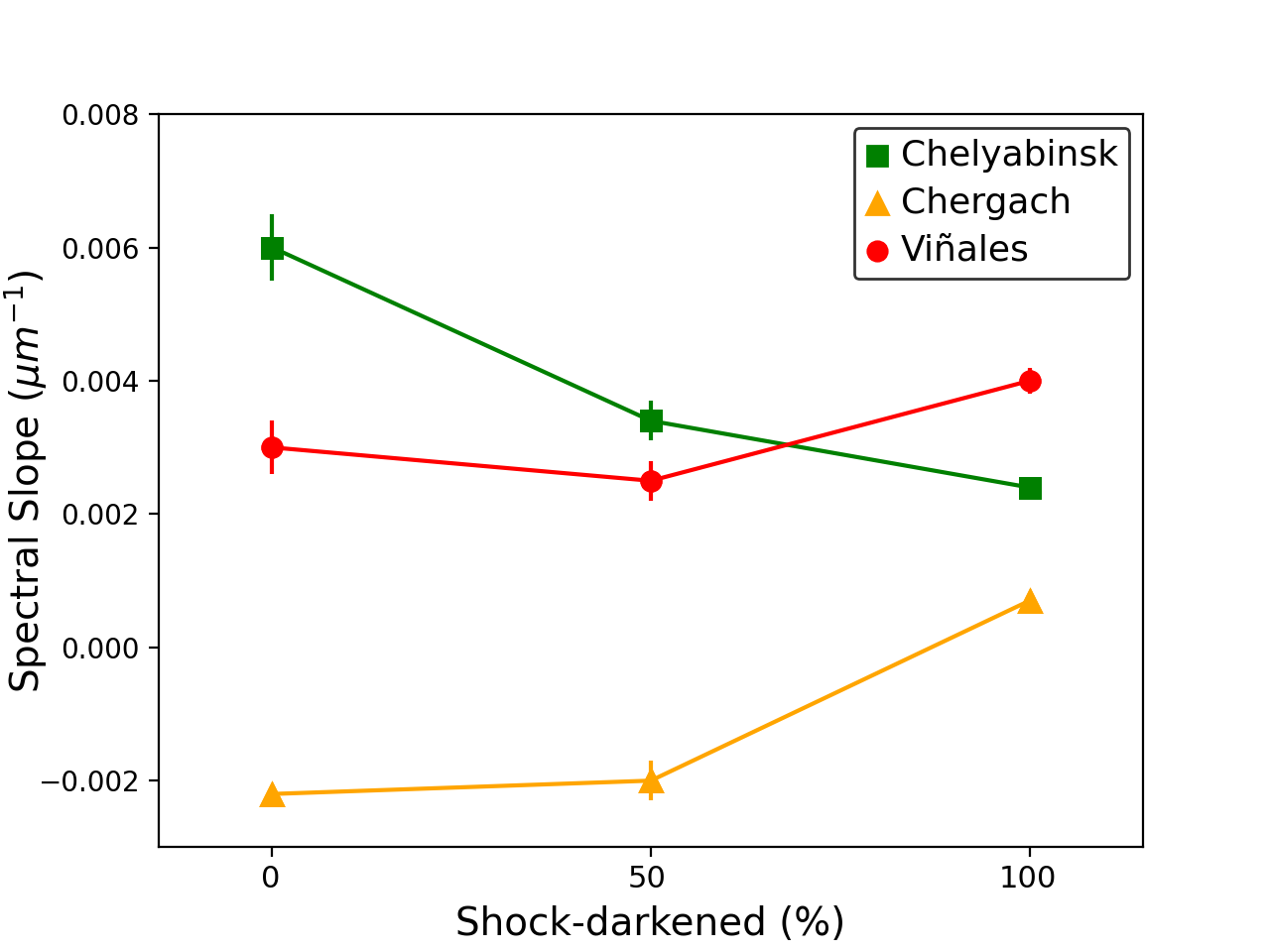}
\includegraphics[width=9cm,angle=0]{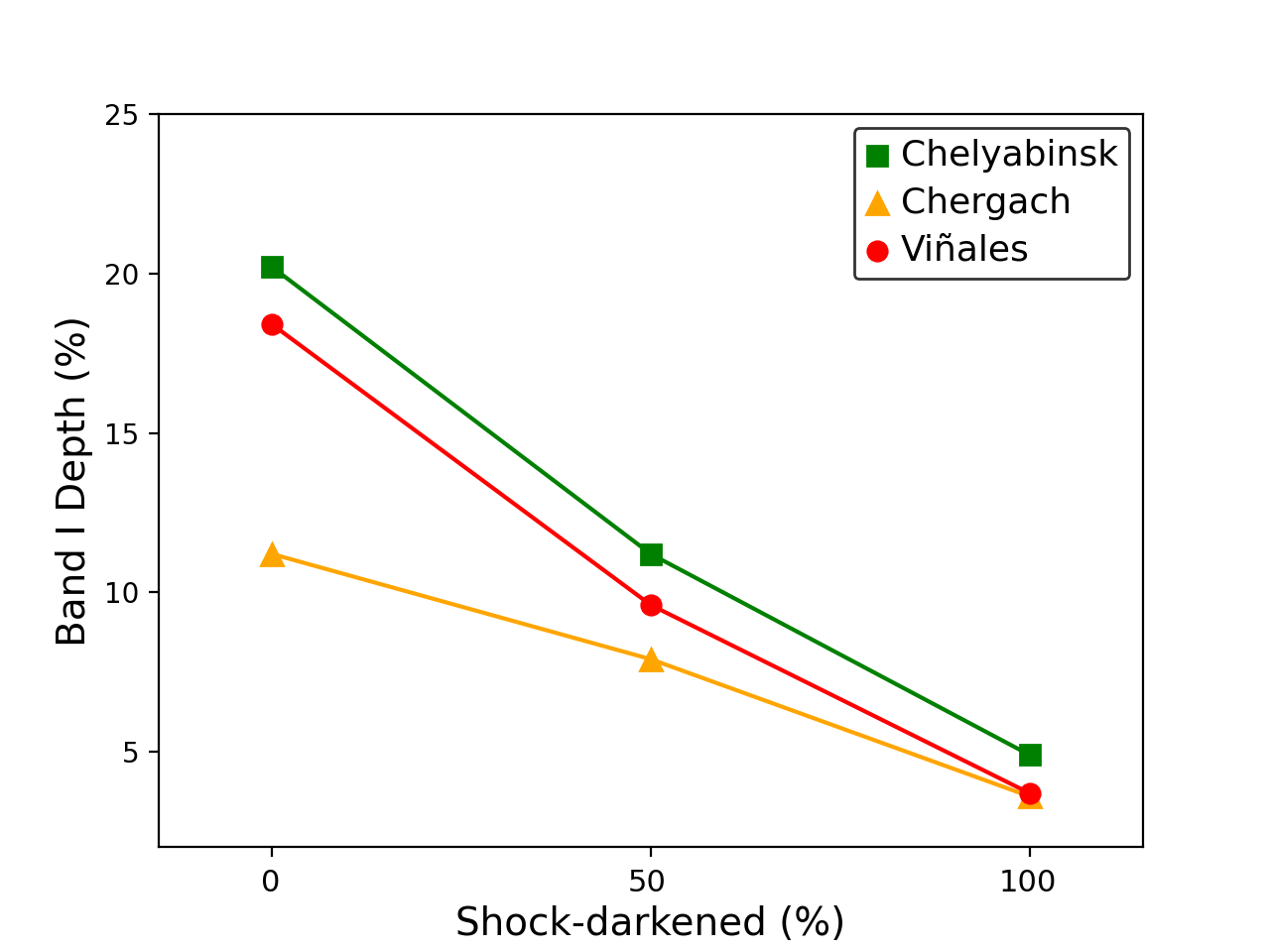}
\hspace{-1.5mm}
\includegraphics[width=9cm,angle=0]{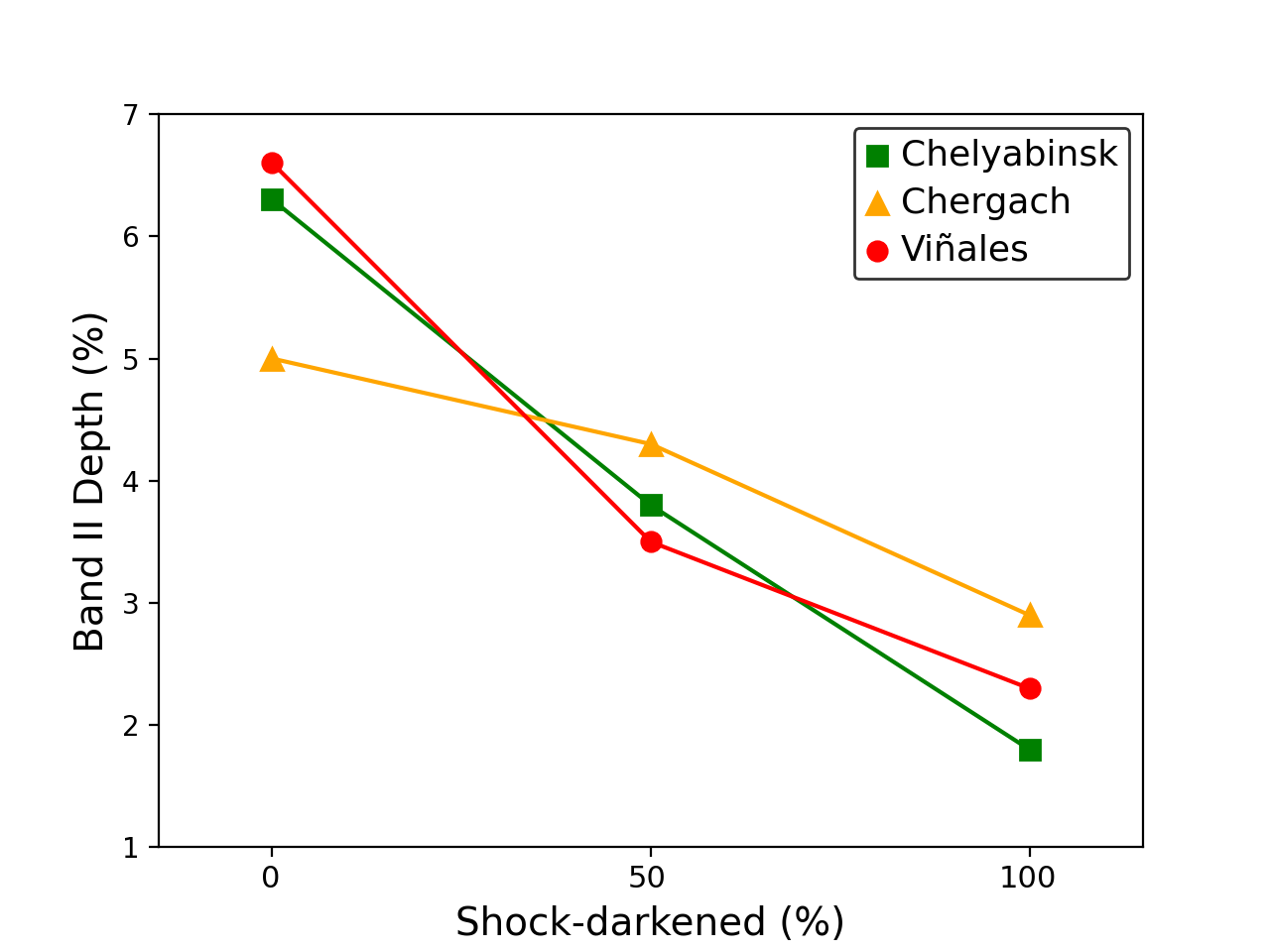}
\caption{Top: Reflectance measured at 0.55 $\mu$m and spectral slope vs. fraction of shock darkening for ordinary chondrites Chelyabinsk (LL5), Chergach (H5), 
and Viñales (L6). 
Bottom: Band I and II depths vs. fraction of shock darkening. Uncertainties are smaller than the symbols.}

\label{f:parameters_SD_OC}
\end{figure}

The spectra of the ordinary chondrites that have both light-colored and dark-colored lithologies (Chergach, Viñales, and Chelyabinsk) are shown in Figures 
\ref{f:Chergach_spec}-\ref{f:Chelyabinsk_spec}. The first thing that we notice in all samples is a significant decrease in reflectance when the shock-darkened material 
is added. The albedo of these samples (Figure \ref{f:parameters_SD_OC}), given by the reflectance measured at 
0.55 $\mu$m, went from a mean value of 0.205 (no shock-darkened material) to 0.124 (50\% shock-darkened material) and to 0.077 (100\% shock-darkened material). No clear trend
is observed for the spectral slope (Figure \ref{f:parameters_SD_OC}). Chelyabinsk spectra show a decrease in spectral slope as the amount of shock darkening increases. The slope of 
Chergach slightly increases from 0 to 50\% shock darkening, but then shows a major increase from 50 to 100\% shock darkening. For Viñales, there is a small decrease in spectral slope from 
0 to 50\% shock darkening and then, like Chergach, there is a more pronounced increase in slope from 50 to 100\% shock darkening. \cite{1994GeCoA..58.3905B} found that 
black chondrites often exhibit a modest red continuum slope compared to normal ordinary chondrites. The inconsistent behavior of the spectral slope in our samples could be the result of a compositional heterogeneity, which can also affect this parameter. In the case of the band depths, there is a consistent trend among the three samples where the band depths decrease with increasing shock darkening (Figure \ref{f:parameters_SD_OC}).

\begin{deluxetable*}{cccccc}

\tablecaption{\label{t:Table2}{\small Taxonomic types for the different meteorite samples. The columns in this table are: name, type, fraction of shock-darkened material 
(SD), principal components PC1' and PC2' and taxonomy.}}

\tablehead{Name&Type&SD ($\%$)&PC1'&PC2'&Taxonomy \\ }

\startdata
Chergach&H5&0&-0.4522&0.1072&Sq \\
Chergach&H5&50&-0.5246&0.0793&C, Ch, Xk, Xn \\
Chergach&H5&100&-0.5162&0.0122&C, Ch, Xk, Xn \\
Tassédet&H5&100&-0.5025&0.0461&C, Ch, Xk, Xn \\
NWA 4860&L4&100&-0.3605&0.0158&C, Ch, Xc, Xe, Xk, Xn \\
Pampa-C&L4&100&-0.1584&-0.0780&K, Xe \\
Ghubara&L5&100&0.3877&-0.3523&L, Xe \\
Tsarev&L5&100&-0.5437&0.0576&C, Ch, Xk, Xn \\
Chico&L6&0&-0.5934&0.3015&Q \\
Renfrow&L6&100&0.1663&-0.0906&S \\
Viñales&L6&0&-0.7752&0.2764&Q \\
Viñales&L6&50&-0.5722&0.0926&C, Ch, Xk, Xn \\
Viñales&L6&100&-0.3738&-0.0409&C, Ch, Xc, Xe, Xk, Xn \\
Chelyabinsk&LL5&0&-0.7464&0.2766&Q \\
Chelyabinsk&LL5&50&-0.6770&0.1390&Q \\
Chelyabinsk&LL5&100&-0.5742&0.0279&C, Cb \\
NWA 7266&Eucrite&0&-0.3794&1.0233&V \\
NWA 7266&Eucrite&50&-0.4743&0.6118&O, Q \\
NWA 7266&Eucrite&100&-0.4899&0.3872&Q \\
NWA 8563&Eucrite&0&-0.1876&0.5774&V \\
NWA 8563$^{*}$&Eucrite&-&-0.1601&0.5043&R \\
NWA 4664$^{*}$&Diogenite&-&-0.2688&0.6187&V \\
\enddata
\tablenotetext{*}{These samples contain light clasts intimately mixed with dark material.}
\end{deluxetable*}

\begin{figure*}[!ht]
\begin{center}
\includegraphics[height=9cm]{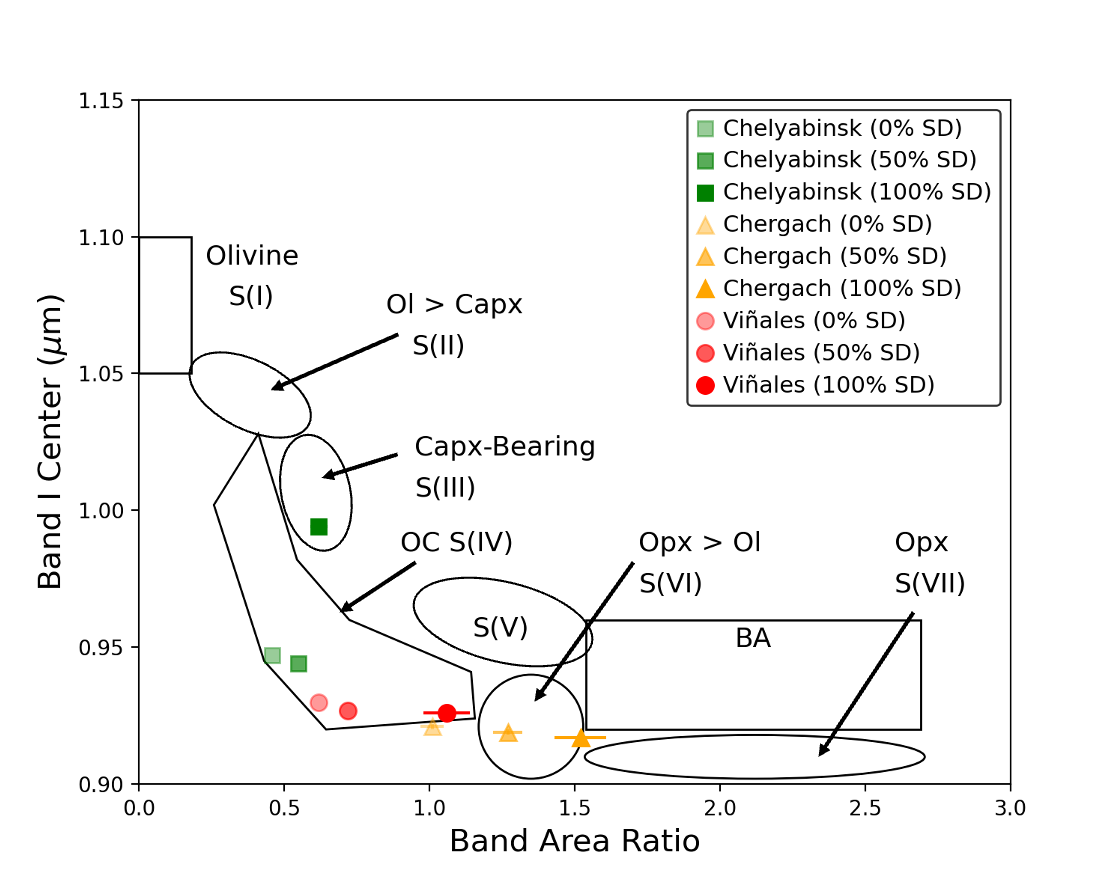}

\caption{\label{f:BIC_OC}{\small Band I center vs. BAR diagram from \cite{1993Icar..106..573G}. Values measured for ordinary chondrites Chelyabinsk (LL5), 
Chergach (H5), and Viñales (L6) 
corresponding to 0\%, 50\%, and 100\% shock darkening (SD) are shown. The increase in SD is represented as an increase in the opacity of the symbols. 
The polygonal region corresponds to the S(IV) subtype associated with ordinary chondrites (OC). The rectangular zone (BA) includes the pyroxene dominated basaltic achondrite assemblages \citep{1993Icar..106..573G}.}}

\end{center}
\end{figure*}

The Band I centers of Chergach and Viñales show little change; however, for Chelyabinsk, there is a significant shift in the Band I center toward longer wavelengths from 
50\% to 100\% shock darkening (Figure \ref{f:BIC_OC}). This shift in the Band I center occurs because of a distortion of the absorption band. In the spectrum corresponding 
to the light-colored lithology, there is one deep central absorption band at $\sim$0.947 $\mu$m, and two right-side lobes at $\sim$1.05 $\mu$m and 
1.17 $\mu$m (Figure \ref{f:Chelyabinsk_spec}). As the amount of shock darkening increases, the absorption band becomes shallower, and the two lobes become weaker, reaching a point at 
100\% shock darkening where the lobe at $\sim$1.05 $\mu$m essentially disappears and merges with the central absorption band, shifting the Band I center to longer wavelengths. This 
does not happen with the Chergach and Viñales spectra because the lobe at $\sim$1.05 $\mu$m is almost absent from the beginning, perhaps due to a lower olivine content in the samples. The 
Band II centers of Chergach, Viñales, and Chelyabinsk were found to shift to longer wavelengths with increasing shock darkening in all cases. This result could be related to 
the suppression of the Band II, which makes it more difficult to obtain an accurate measurement of the band center.

We also found that the BAR of Chergach, Viñales, and Chelyabinsk increased when the shock-darkened material was added. Figure \ref{f:BIC_OC} shows the seven 
compositional subtypes in which S-type asteroids can be classified in the Band I center vs. BAR plot of \cite{1993Icar..106..573G}. In this Figure, Chergach moved from the S(IV) to the S(VI) subtype when the amount of shock-darkened material increased to 50\%, and it almost reached the S(VII) subtype for the sample with 100\% shock-darkened 
material. Viñales stayed within the S(IV) subtype, but it moved from one extreme to the other in the ordinary chondrite region. In the 
case of Chelyabinsk, the increase in BAR was smaller; however, because there was also a shift in the Band I center, it moved from the S(IV) subtype to the S(III) subtype (Figure \ref{f:BIC_OC}).

\begin{figure*}[!ht]
\begin{center}
\includegraphics[height=8.5cm]{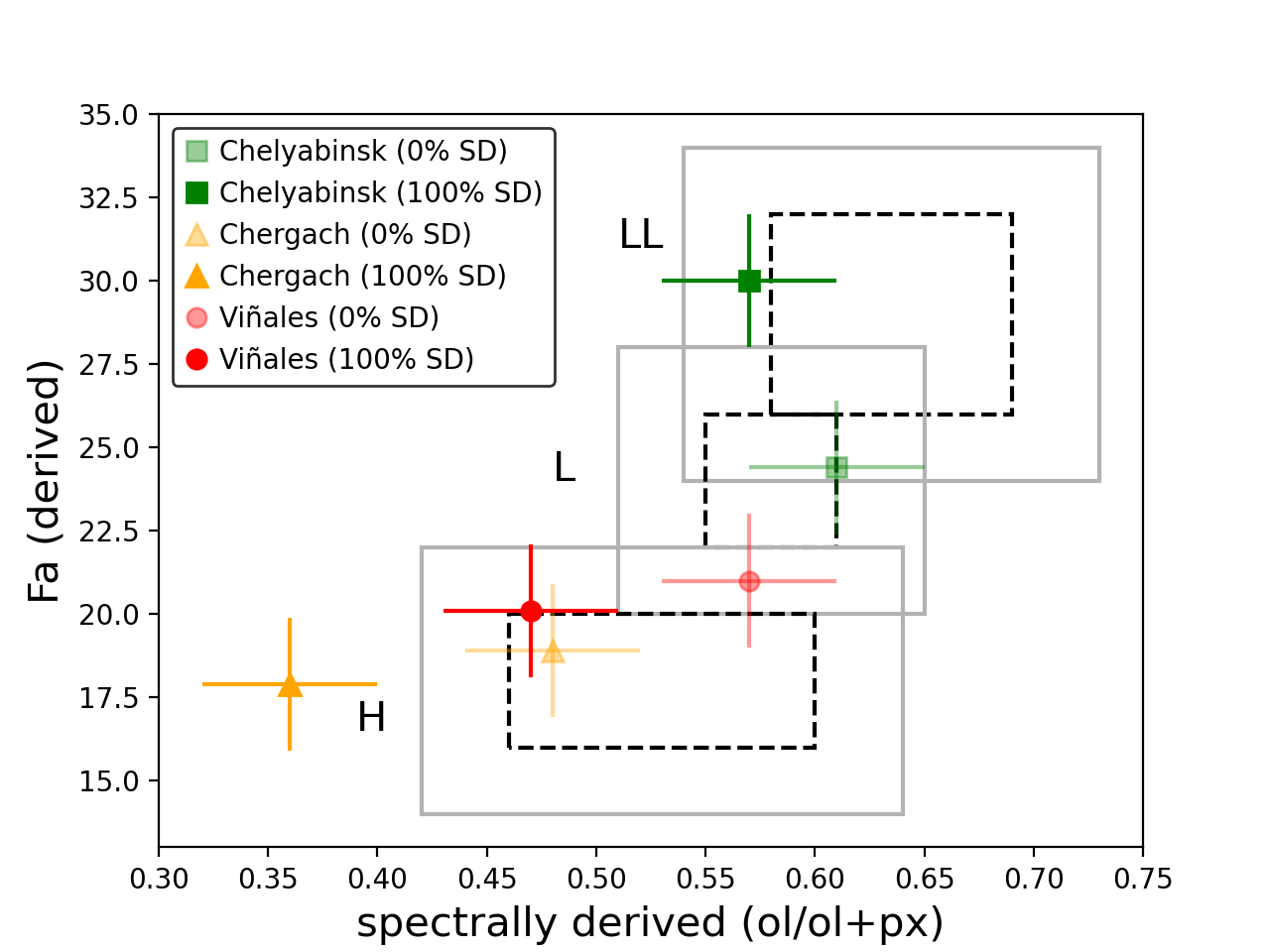}

\caption{\label{f:Fa_olopx}{\small Molar content of Fa vs. ol/(ol+px) ratio for ordinary chondrites Chelyabinsk (LL5), Chergach (H5), and Viñales (L6) corresponding to 0\% and 100\% shock 
darkening (SD). The increase in SD is represented as an increase in the opacity of the symbols. The error bars correspond to the uncertainties derived by \cite{2020AJ....159..146S}, 2.0 mol\%
for Fa and 0.04 for the ol/(ol+px) ratio. Black dashed boxes represent the range of measured values for each ordinary chondrite subgroup. Gray solid boxes correspond to the uncertainties
associated with the spectrally derived values. Figure adapted from \cite{2010Icar..208..789D}.}}

\end{center}
\end{figure*}

The Band I center and BAR are used to determine the composition of S-complex asteroids that fall in the S(IV) subtype region \citep[e.g.,][]{2010Icar..208..789D, 2020AJ....159..146S}. We used the equation of \cite{2020AJ....159..146S} to determine if the variations in these band parameters due to shock darkening could affect 
the derived composition. The Band I center was used to determine the olivine and pyroxene chemistries, which are given by the mol\% of Fa and Fs, respectively. 
The BAR was used to calculate the olivine-to-pyroxene ratio (ol/(ol+px)). Figure \ref{f:Fa_olopx} shows the molar content of Fa vs. ol/(ol+px) for Chergach, Viñales, and 
Chelyabinsk corresponding to 0\% and 100\% shock darkening. In this Figure, we can see that the large increase in BAR for Chergach and Viñales resulted in an underestimation of 
the ol/(ol+px) that is larger than the uncertainty. The ol/(ol+px) for Chergach decreased from 0.48 to 0.36 and it moved out of the 
H-chondrites region. For Viñales, the ol/(ol+px) decreased from 0.57 to 0.47, moving Viñales from the L-chondrites region to the H-chondrites region. For both Chergach and Viñales, 
there is little change in the Fa value, which is expected since the Band I center of these samples did not change much. The ol/(ol+px) of Chelyabinsk was the least affected, decreasing from 0.61 to 0.57, a change that is of the order of the uncertainty. Due to the shift in the Band I center, the Fa value calculated for Chelyabinsk increased from 24.4\% to 
30.0\%.

\begin{figure*}[!ht]
\begin{center}
\includegraphics[height=8.5cm]{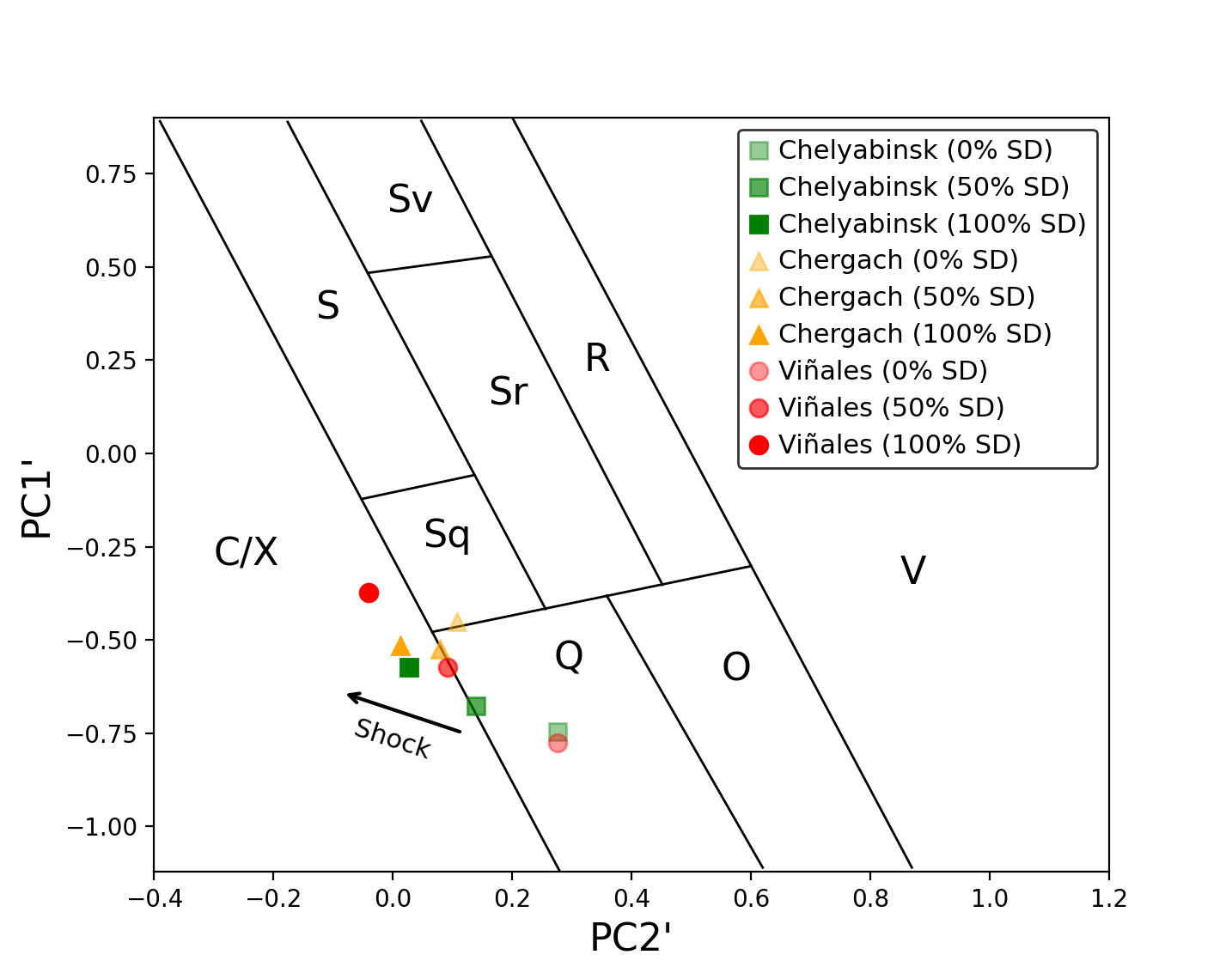}

\caption{\label{f:OC_PC_1}{\small PC2' vs. PC1' diagram from \cite{2009Icar..202..160D}. PC values calculated for ordinary chondrites Chelyabinsk (LL5), 
Chergach (H5), and Viñales (L6) corresponding to 0\%, 50\%, and 100\% shock darkening (SD) are depicted with different symbols. The increase in SD is represented as an increase in the opacity of the symbols. The arrow
indicates the direction in which shock darkening increases.}}

\end{center}
\end{figure*}

The results of the taxonomic classification for Chergach, Viñales, and Chelyabinsk are shown in Figure \ref{f:OC_PC_1}. For both Viñales and Chelyabinsk, the spectra 
corresponding to the light-colored lithology were classified as Q-types. Adding 50\% of shock-darkened material moves these samples to the line $\alpha$, which separates the 
S-complex from the C- and X-complexes. At this point, Chelyabinsk is still classified as a Q-type, but Viñales is being classified as C/X-complex. Increasing to 100\% shock-darkened 
material moves Viñales and Chelyabinsk further into the C/X-complex region, and Chelyabinsk is now classified as a C- or Cb-type. Chergach is initially classified as Sq-type 
(0\% shock darkening), and adding 50\% of shock-darkened material is enough to change the taxonomic classification to C/X-complex. These results are consistent with the 
findings of \cite{2014Icar..237..116R}.

\begin{figure} 
\includegraphics[width=9.5cm,angle=0]{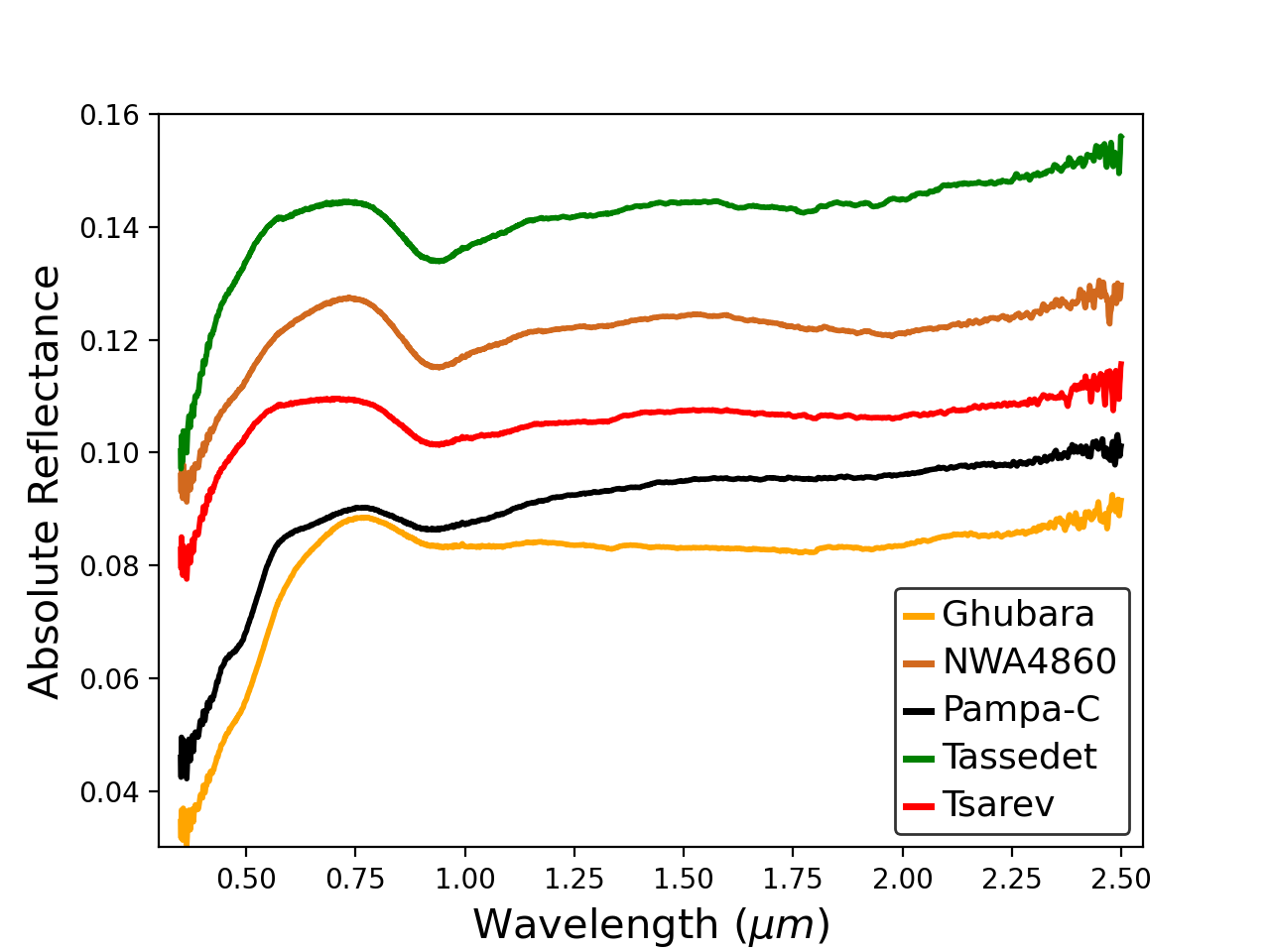}
\hspace{-5mm}
\includegraphics[width=9.5cm,angle=0]{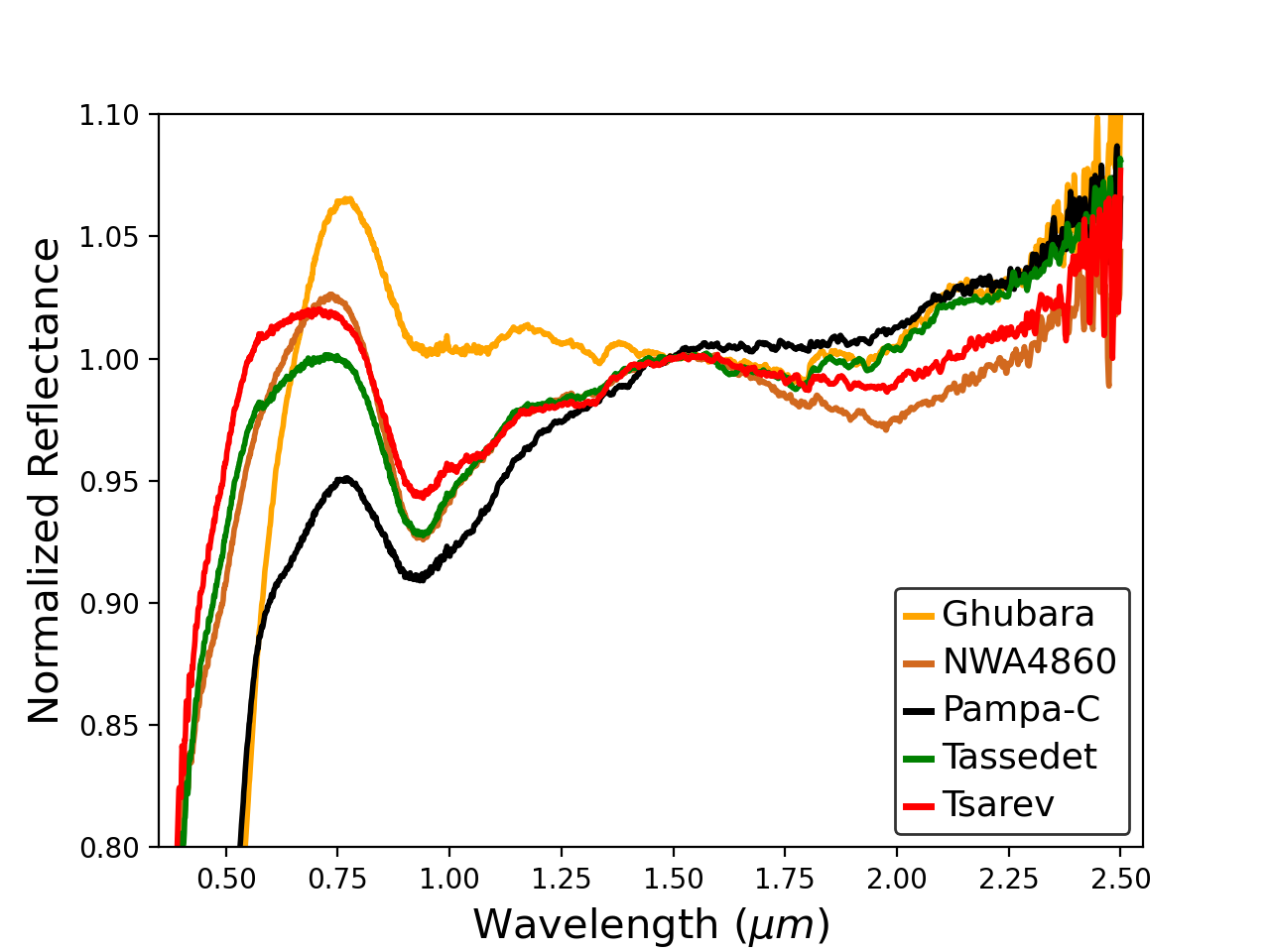}
\caption{Left: Spectra of the L5 Ghubara, L4 NWA 4860, L4 Pampa-C, H5 Tassédet, and L5 Tsarev. Right: The same spectra normalized to unity at 1.5 $\mu$m.}

\label{f:OC_100SD}
\end{figure}

\begin{figure} 
\includegraphics[width=9.5cm,angle=0]{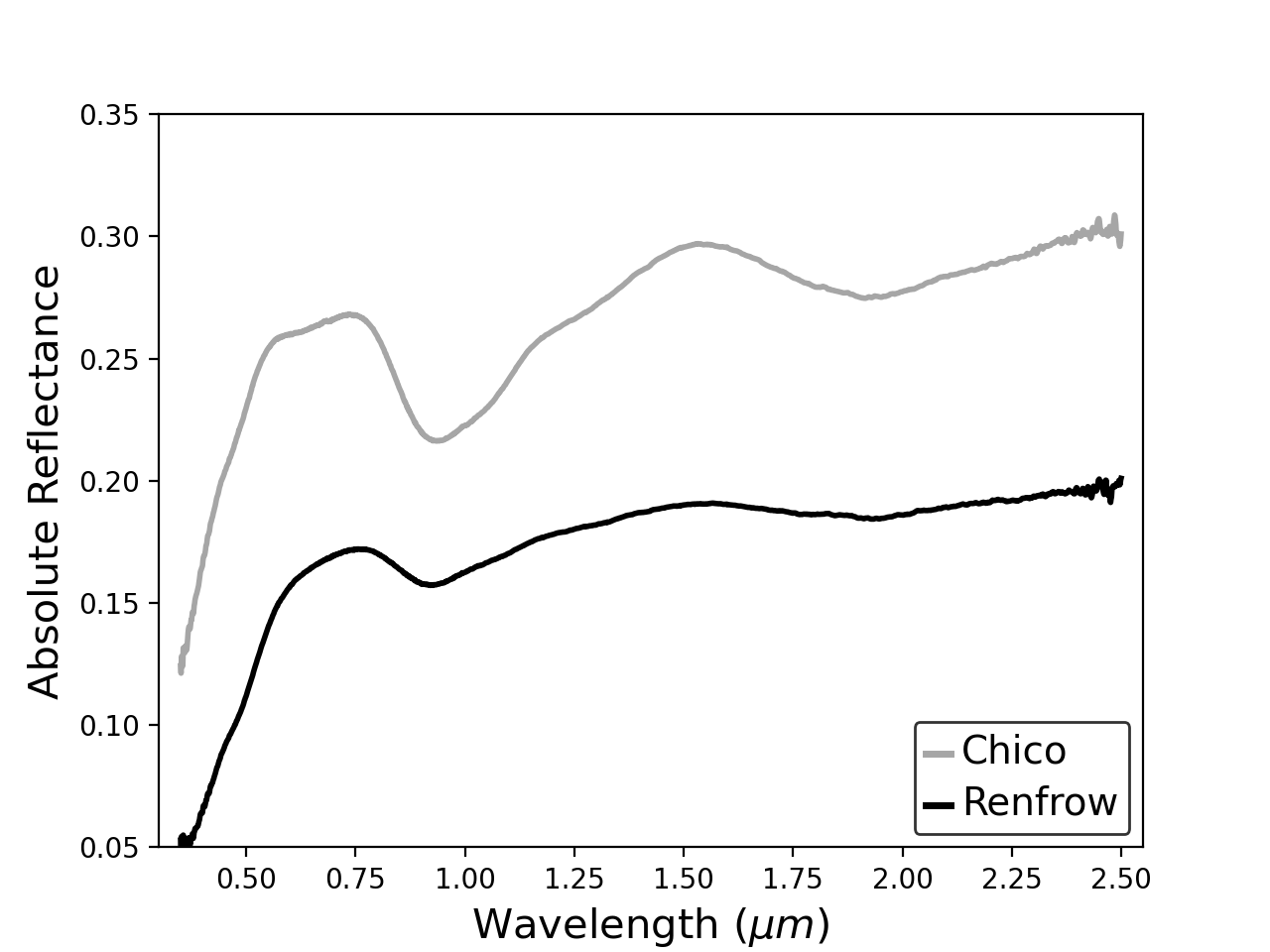}
\hspace{-5mm}
\includegraphics[width=9.5cm,angle=0]{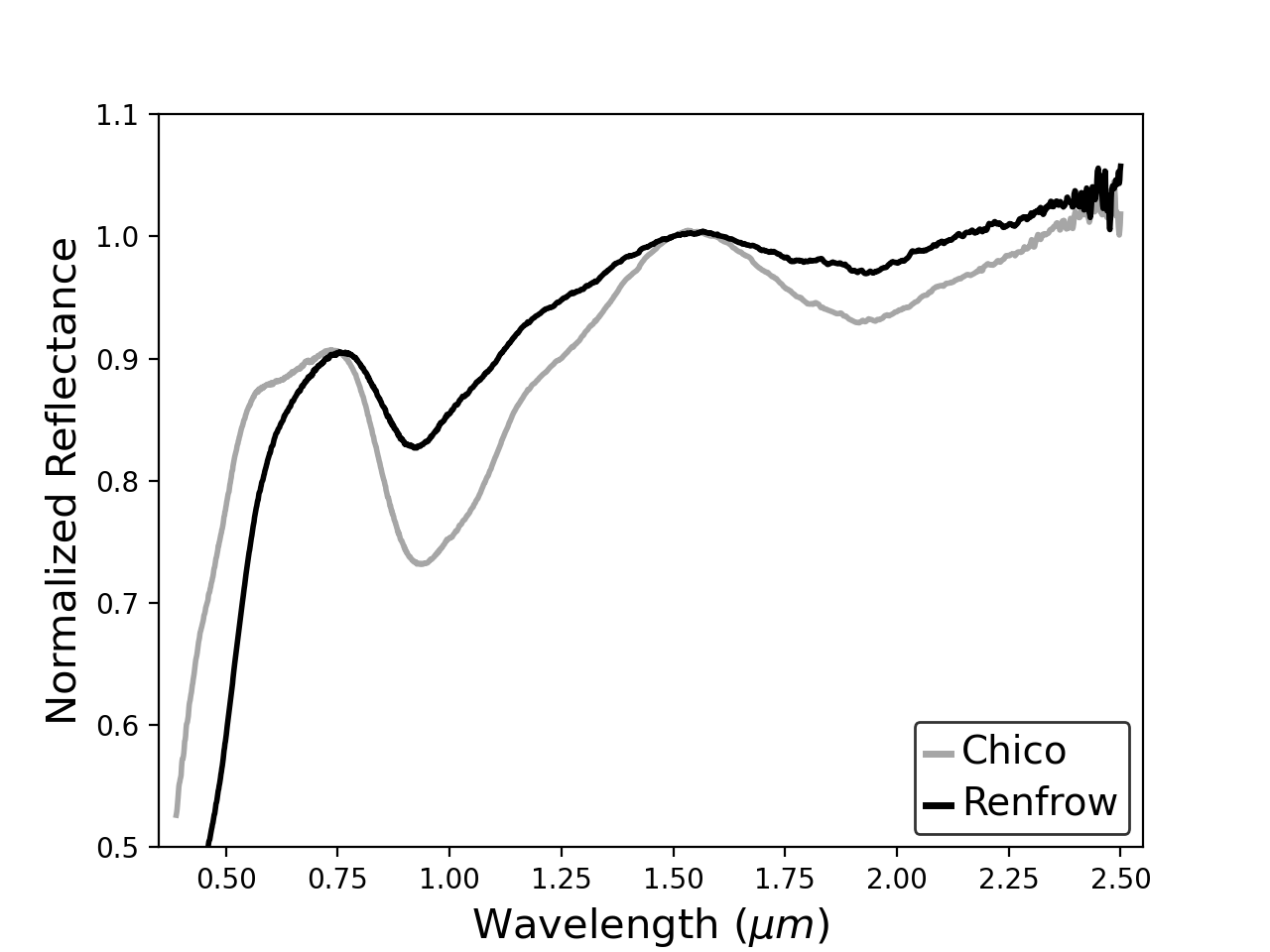}
\caption{Left: Spectra of the L6 ordinary chondrites Chico and Renfrow. Right: The same spectra normalized to unity at 1.5 $\mu$m.}

\label{f:Chico_Renfrow}
\end{figure}

\begin{figure*}[!ht]
\begin{center}
\includegraphics[height=9cm]{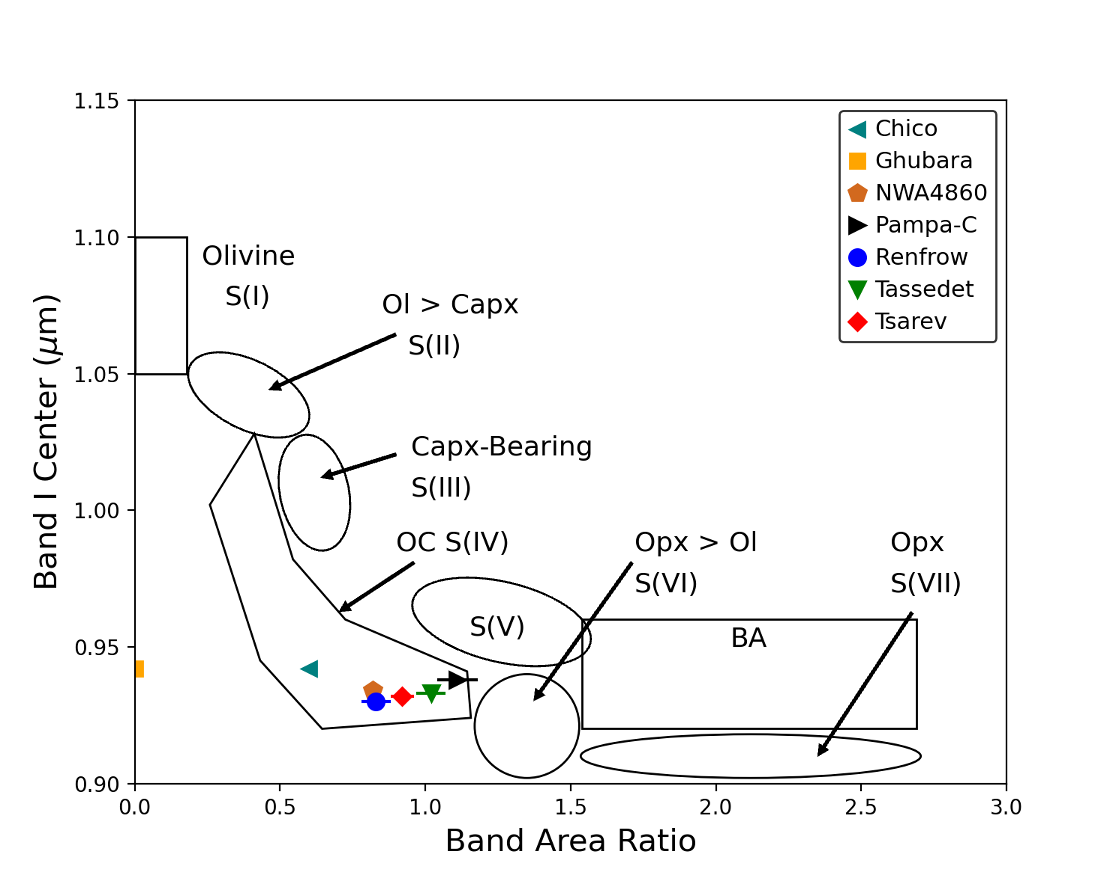}

\caption{\label{f:BIC_OCB}{\small Band I center vs. BAR diagram from \cite{1993Icar..106..573G}. Values measured for ordinary chondrites Chico (L6), Ghubara (L5), NWA 4860 (L4), 
Pampa-C (L4), Renfrow (L6), Tassédet (H5), and Tsarev (L5) are shown. These samples do not show a light/dark structure, but a rather dark and uniform color. The polygonal 
region corresponds to the S(IV) subtype associated with ordinary chondrites (OC). The rectangular zone (BA) includes the pyroxene dominated basaltic achondrite assemblages \citep{1993Icar..106..573G}.}}

\end{center}
\end{figure*}

\begin{figure*}[!ht]
\begin{center}
\includegraphics[height=9cm]{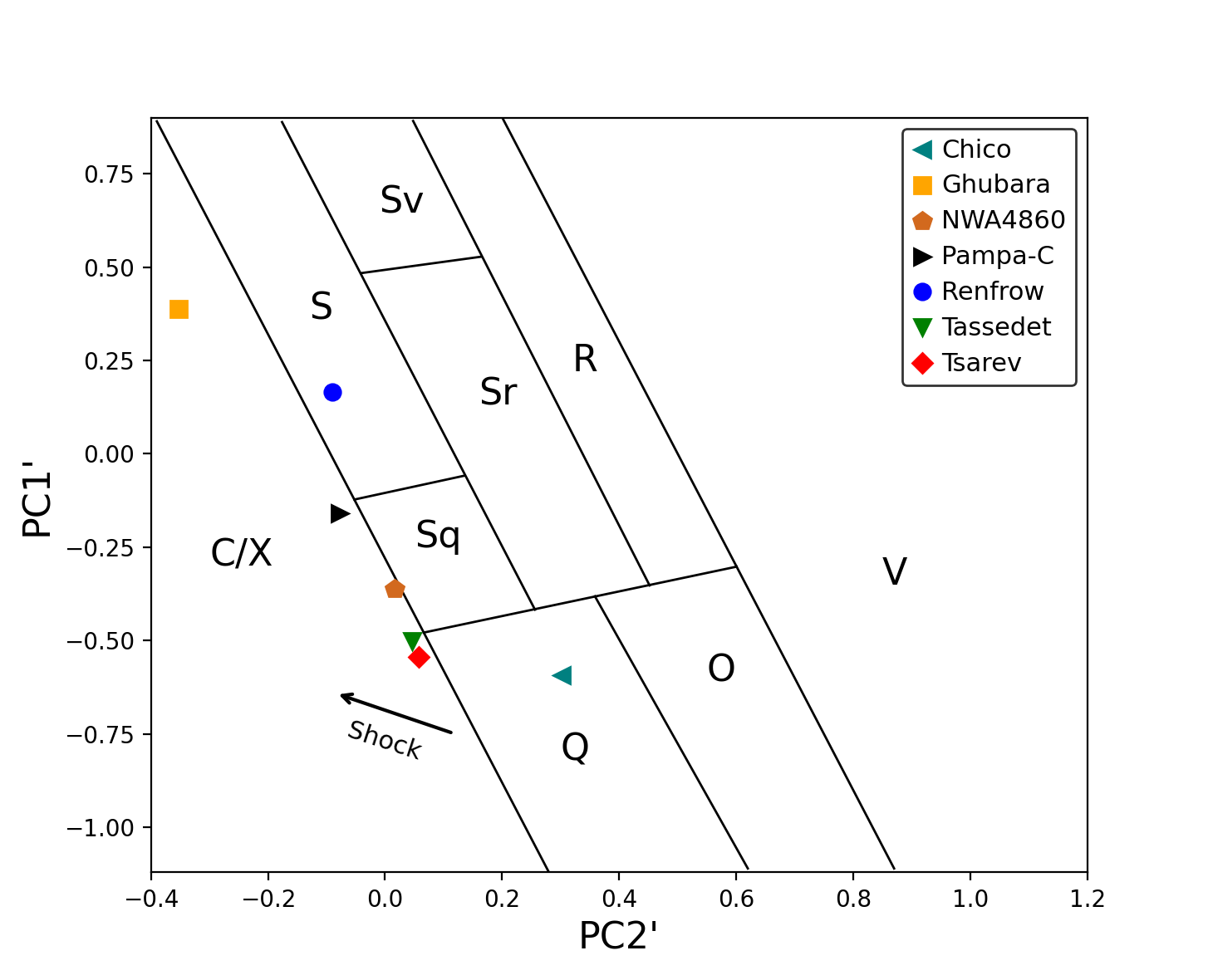}

\caption{\label{f:OC_PC_2}{\small PC2' vs. PC1' diagram from \cite{2009Icar..202..160D}. PC values calculated for ordinary chondrites Chico (L6), Ghubara (L5), NWA 4860 (L4), 
Pampa-C (L4), Renfrow (L6), Tassédet (H5), and Tsarev (L5) are depicted with different symbols.}}

\end{center}
\end{figure*}

Ordinary chondrites Tassédet, NWA 4860, Pampa-C, Ghubara, Tsarev, Chico, and Renfrow do not show a light/dark structure, but a rather dark and uniform color. The 
spectra of these meteorites are shown in Figures \ref{f:OC_100SD} and \ref{f:Chico_Renfrow}. The albedo of all these meteorites, with the exception of Chico, 
range from $\sim$0.07 to 0.14, which is roughly the range found for the samples containing 50\%-100\% shock-darkened material. This albedo range is also consistent 
with values reported by previous studies \citep[e.g.,][]{1976JGR....81..905G, 1991Metic..26..279B}. The albedo of Chico (0.254) is the 
highest of all the ordinary chondrites analyzed in this study, which indicates that in this particular sample, there is little or no shock-darkened material present. The band 
depths of the H chondrite Tassédet are similar to those of the H chondrite Chergach with 50\% of shock darkening. Similarly, the L chondrite NWA 4860 has band depths 
with values close to the L chondrite Viñales containing 50\% of shock-darkened material. The Band depth values for the L chondrite Renfrow, on the other hand, fall 
between those measured for Viñales corresponding to 0\%-50\% of shock darkening. For the L chondrites Pampa-C, Ghubara, and Tsarev, we found that the band depths 
are similar to the values measured for the samples of Viñales containing 50-100\% shock-darkened material. In the spectrum of Ghubara, the 2-$\mu$m band is not present 
because the reflectance maximum at $\sim$1.5 $\mu$m has been completely suppressed, giving this spectrum an appearance similar to some carbonaceous chondrites. The 
spectrum of Chico has the deepest absorption bands of all the ordinary chondrites, even more than the samples of Chelyabinsk and Viñales with no shock-darkened material. This is 
also an indication that this sample does not contain any shock-darkened material despite its relatively dull color.

In the Band I center vs. BAR plot (Figure \ref{f:BIC_OCB}), all these meteorites (with the exception of Ghubara) fall inside the S(IV) subtype region associated with ordinary 
chondrites. Since the 2-$\mu$m band in Ghubara is absent, the BAR for this sample is zero. The taxonomic classification of these meteorites (Figure \ref{f:OC_PC_2}) 
showed that Ghubara was classified as L- or Xe-type, Tassédet, NWA 4860, and Tsarev were all assigned ambiguous classifications in the C- and X-complex, and 
Pampa-C was classified as either K- or Xe-type (Table 2). Chico and Renfrow are the only two that were assigned the taxonomic types expected for a typical ordinary 
chondrite, Q- and S-type, respectively. The spectra of these two meteorites are compared in Figure \ref{f:Chico_Renfrow}, where we can see that the spectrum of Renfrow shows a 
much lower reflectance and weaker absorption bands than Chico.

\subsection{Eucrites and Diogenites}
 
\begin{figure}[h!] 
\includegraphics[width=9.5cm,angle=0]{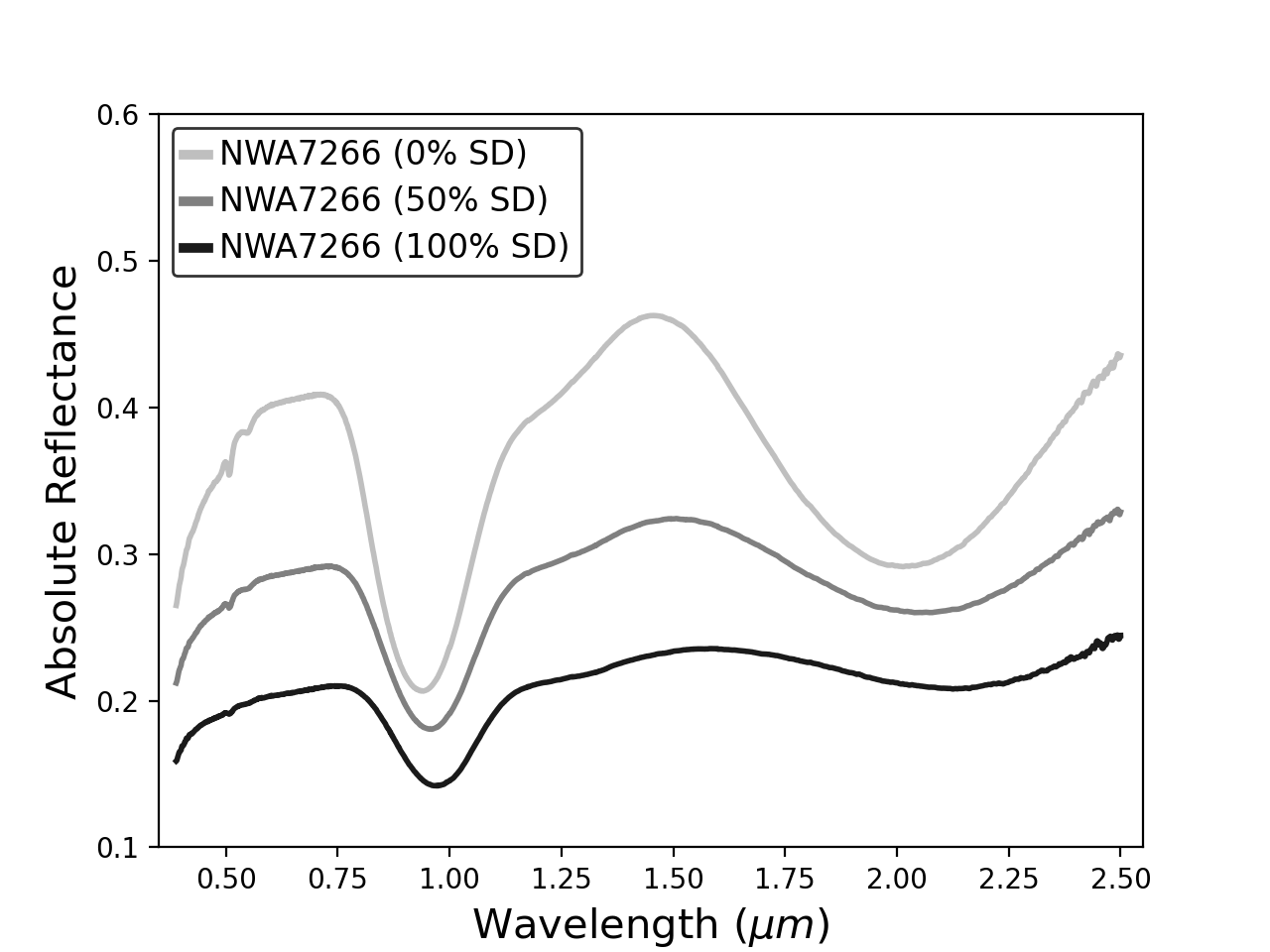}
\hspace{-5mm}
\includegraphics[width=9.5cm,angle=0]{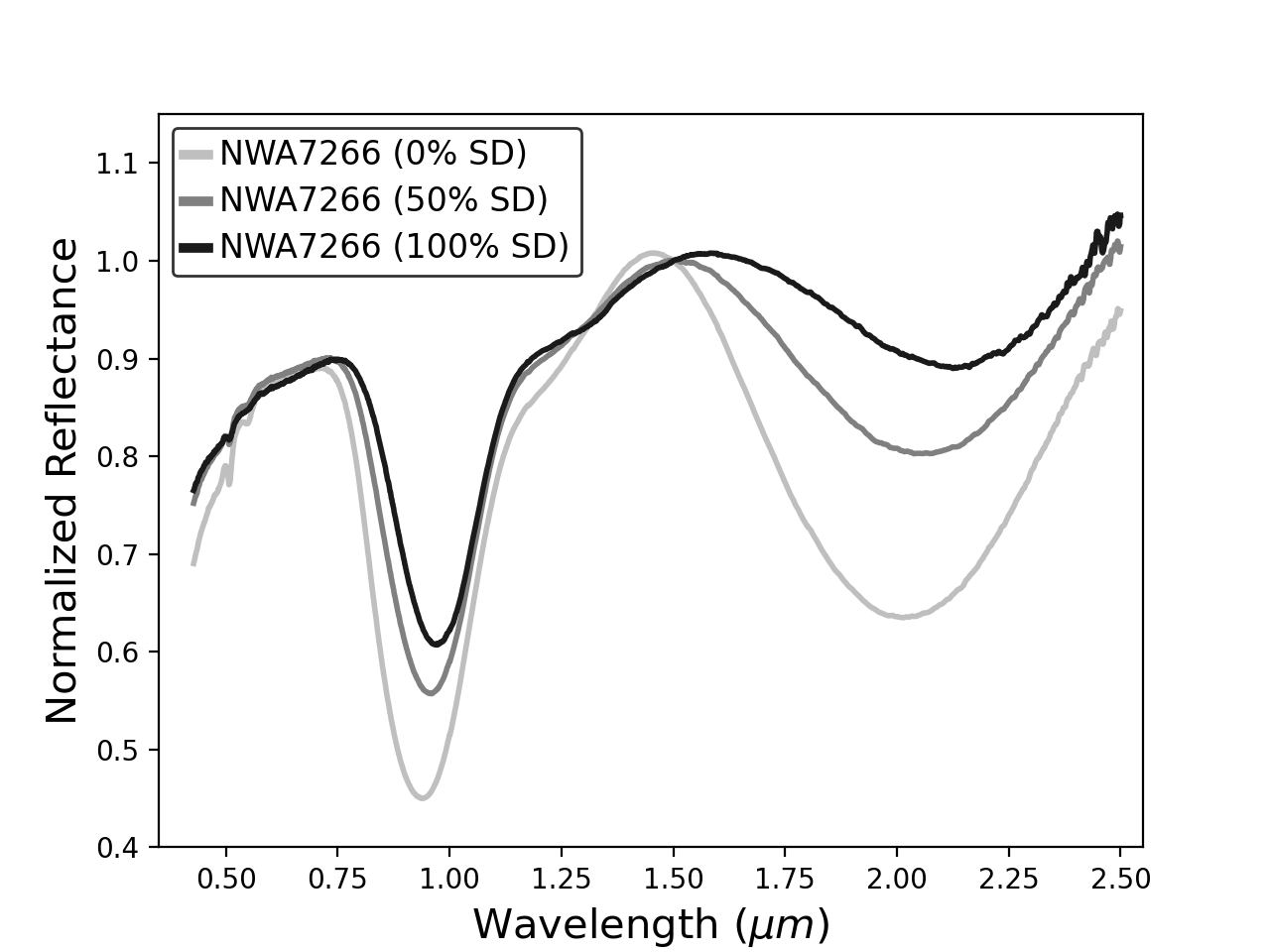}
\caption{Left: Spectra of eucrite NWA 7266 corresponding to 0\%, 50\%, and 100\% shock darkening (SD). Right: The same spectra normalized to unity at 1.5 $\mu$m.}

\label{f:NWA7266_spec}
\end{figure}

\begin{figure} 
\includegraphics[width=9cm,angle=0]{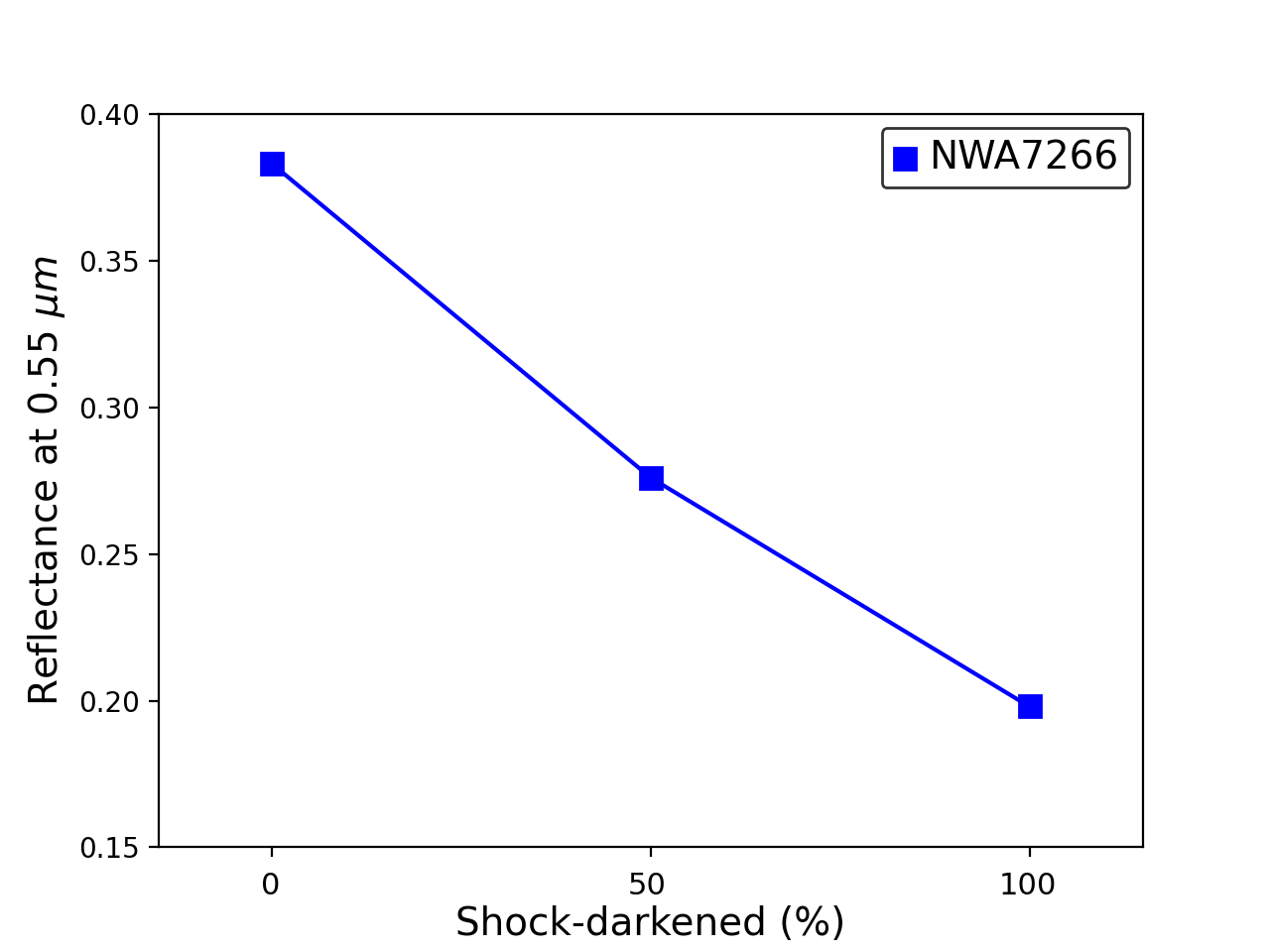}
\hspace{-5mm}
\includegraphics[width=9cm,angle=0]{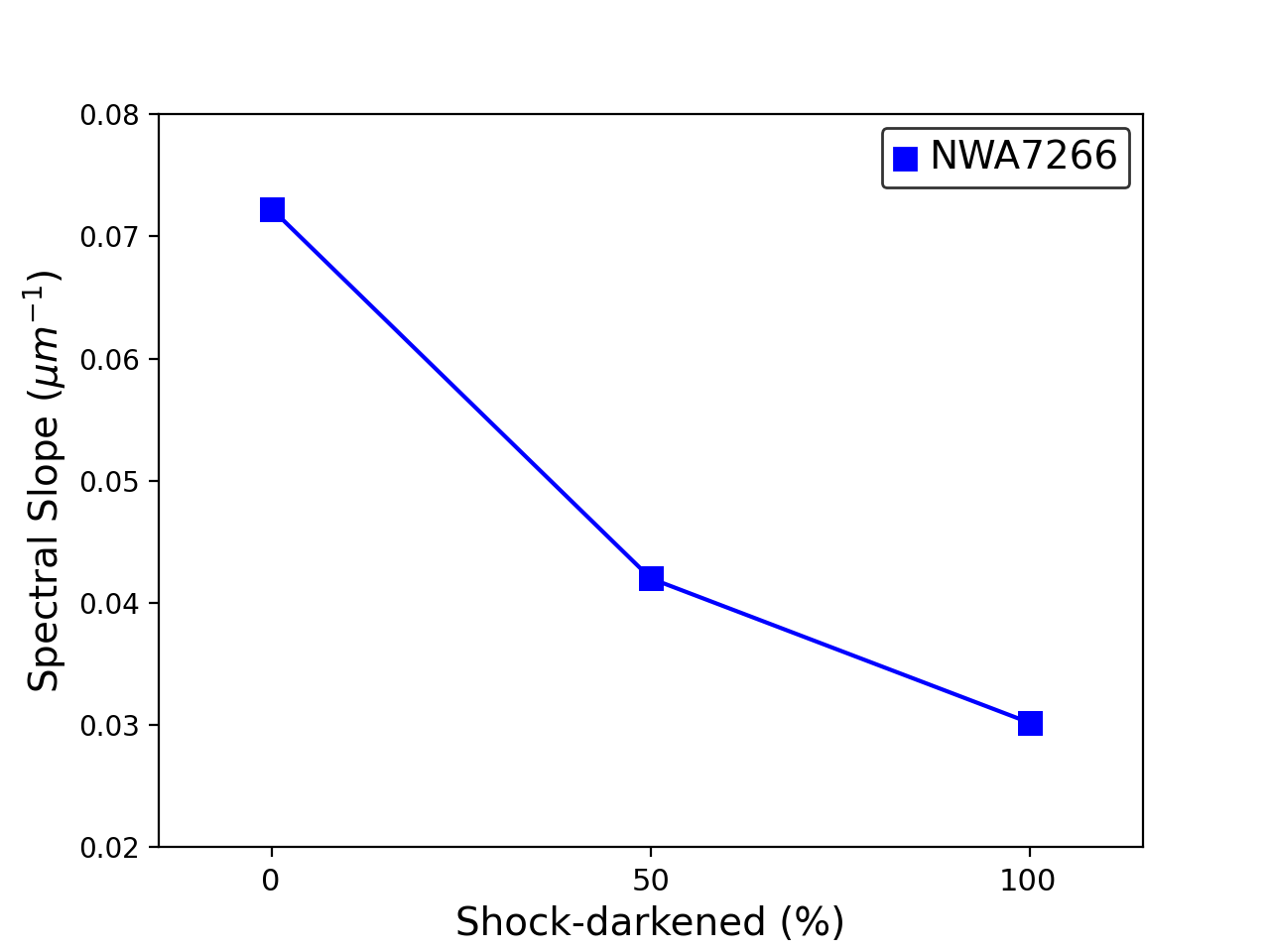}
\includegraphics[width=9cm,angle=0]{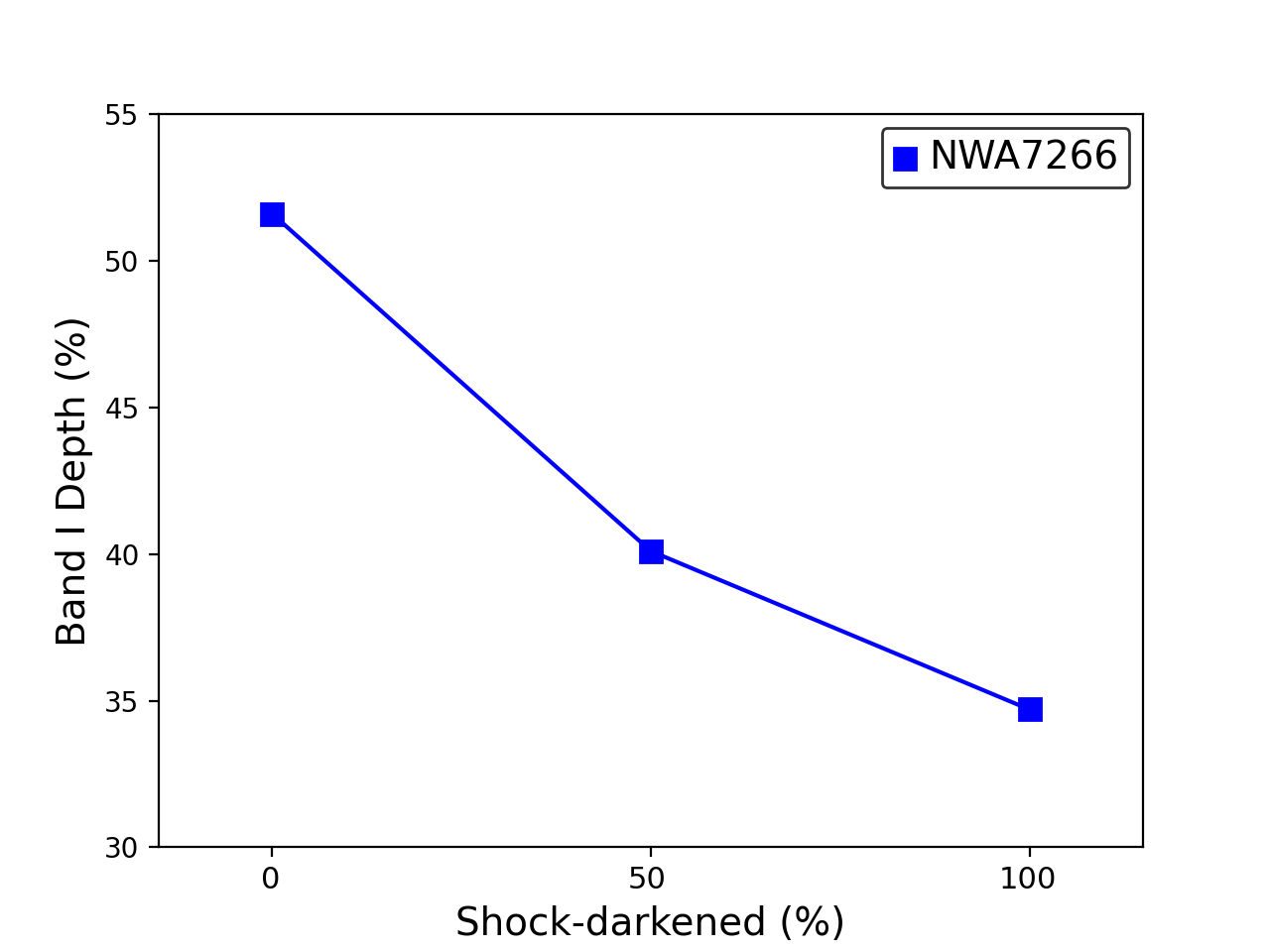}
\hspace{-1.5mm}
\includegraphics[width=9cm,angle=0]{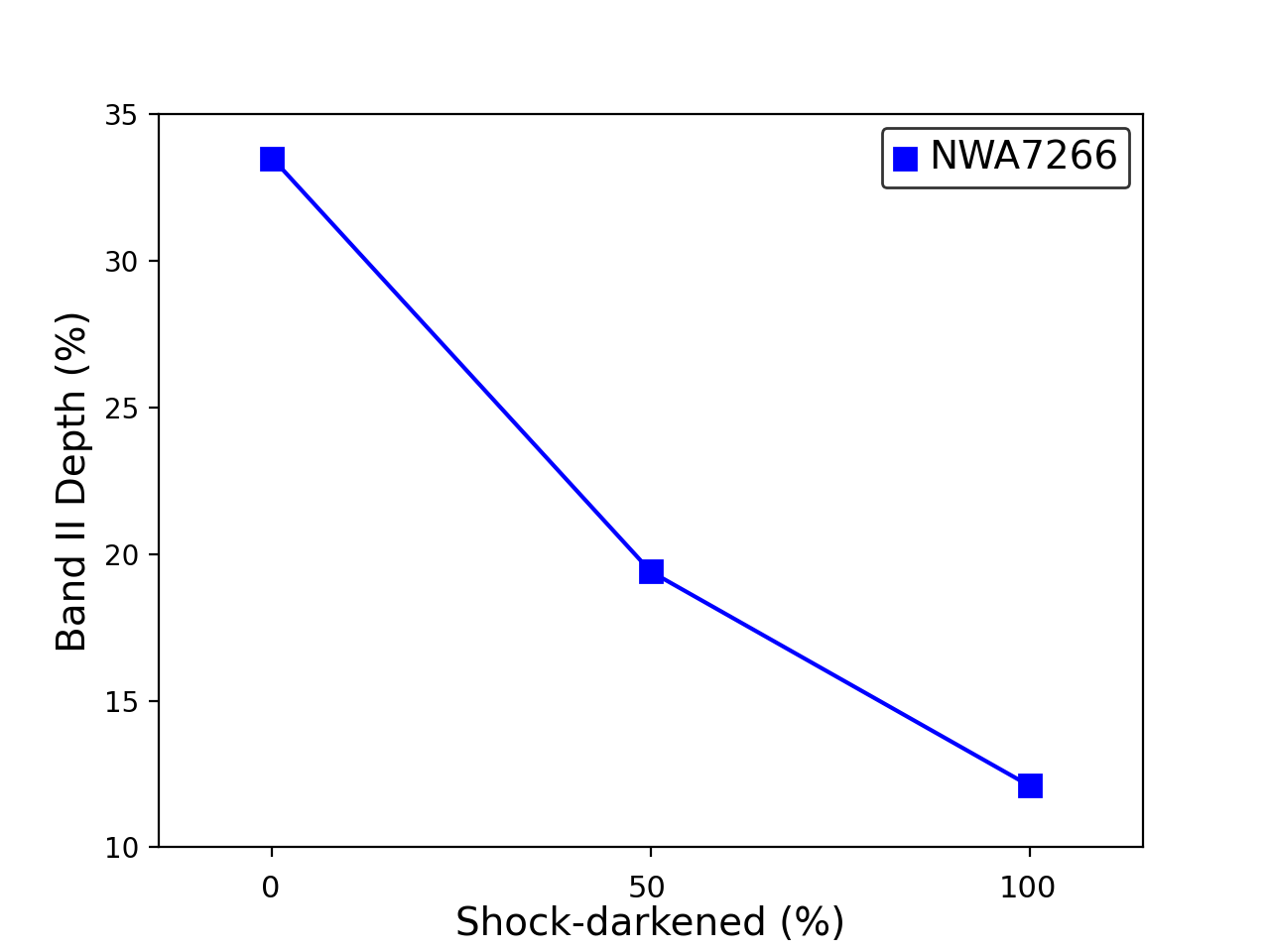}

\caption{Top: Reflectance measured at 0.55 $\mu$m and spectral slope vs. fraction of shock darkening for eucrite NWA 7266. 
Bottom: Band I and II depths vs. fraction of shock darkening. Uncertainties are smaller than the symbols.}
\label{f:parameters_SD_HED}
\end{figure}

Figure \ref{f:NWA7266_spec} shows the spectra obtained for eucrite NWA 7266. Similarly to the ordinary chondrites, there is a significant decrease in reflectance with increasing 
shock darkening for this meteorite. The albedo of this sample (Figure \ref{f:parameters_SD_HED}) 
went from 0.383 (0\% shock darkening) to 0.276 (50\% shock darkening) and to 0.198 (100\% shock darkening). In terms of spectral slope, we observed that 
NWA 7266 spectra behave similarly to Chelyabinsk spectra, i.e., there is a decrease in spectral slope as the amount of shock darkening increases 
(Figure \ref{f:parameters_SD_HED}). We also observed a suppression of the absorption bands with increasing shock darkening similar to the ordinary chondrites 
(Figure \ref{f:parameters_SD_HED}). 

\begin{figure*}[h]
\begin{center}
\includegraphics[height=8.5cm]{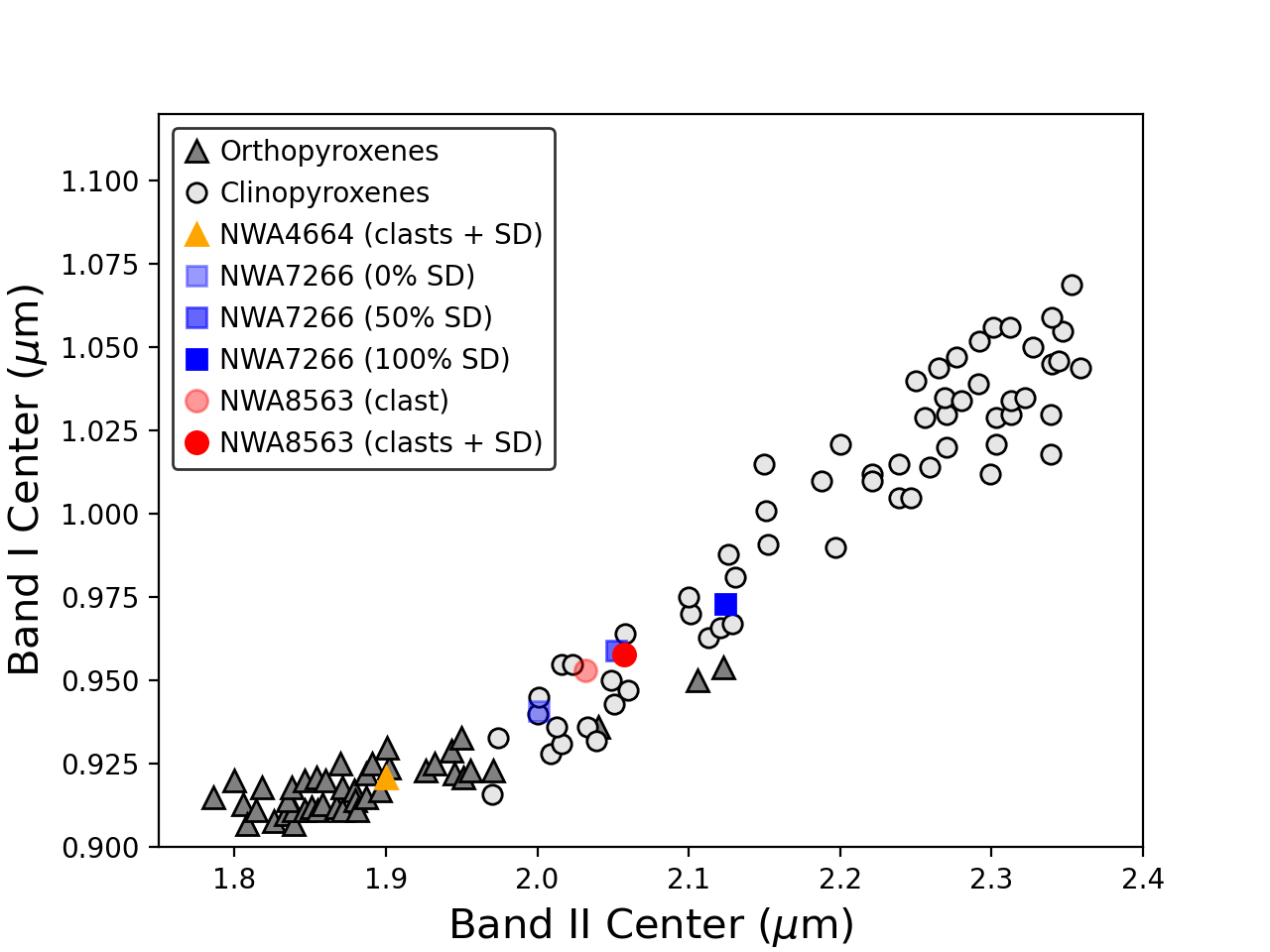}

\caption{\label{f:BIC_BIIC}{\small Band I center vs. Band II center for diogenite NWA 4664 and eucrites NWA 7266 and NWA 8563. The increase in shock darkening (SD) is 
represented as an increase in the opacity of the symbols. Measured band centers for orthopyroxenes ($<$11\% Wo) and clinopyroxenes ($>$11\% Wo) from \cite{1974JGR....79.4829A} and 
\cite{1991JGR....9622809C} are also shown.}}

\end{center}
\end{figure*}

The Band I and II centers of NWA 7266 shift to longer wavelengths as the amount of shock-darkened material increases (Figure \ref{f:BIC_BIIC}). The Band I center was found to be 
0.941 $\mu$m (0\% shock darkening) and shifted to 0.973 $\mu$m (100\% shock darkening), while the Band II center shifted from 2.001 $\mu$m to 2.124 $\mu$m. Unlike the 
ordinary chondrites, the absorption bands of the spectrum corresponding to 100\% shock-darkened material are still deep enough to easily measure the band centers. Therefore, the 
variations in the band centers are not related to a distortion of the absorption bands or a low S/N. The Band I center in low-Ca pyroxenes is typically in the range 
of 0.90-0.93 $\mu$m and the Band II center between $\sim$1.80-2.10 $\mu$m. High-Ca Type B clinopyroxene, on the other hand, show absorption bands centered at 
$\sim$0.91-1.06 $\mu$m and 1.97-2.35 $\mu$m. The band centers of NWA 7266 are consistent with the presence of high-Ca clinopyroxene, and the shift to longer 
wavelengths could be the result of a heterogeneous sample. 

\begin{figure*}
\begin{center}
\includegraphics[height=9cm]{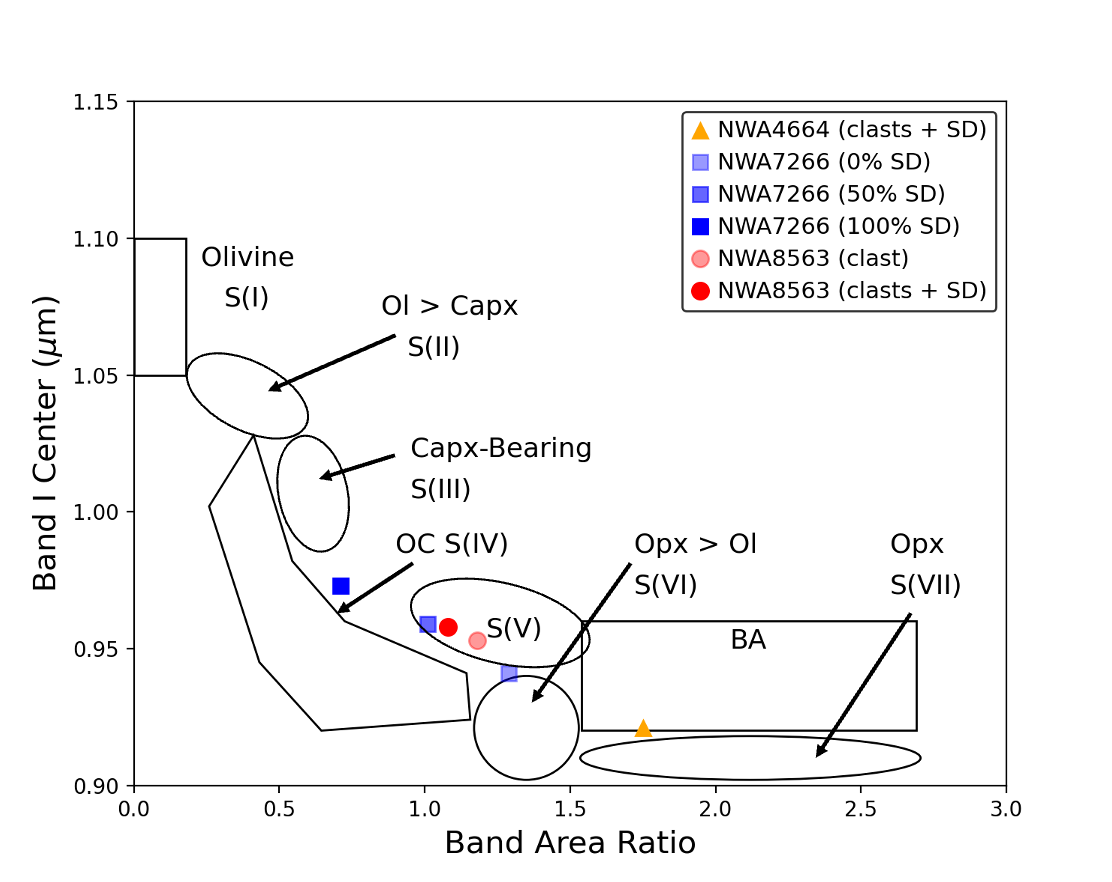}

\caption{\label{f:BIC_HED}{\small  Band I center vs. BAR diagram from \cite{1993Icar..106..573G}. Values measured for diogenite NWA 4664 and eucrites NWA 7266 and NWA 
8563 corresponding to different fractions of shock darkening (SD) are shown. The increase in SD is represented as an increase in the opacity of the symbols. The polygonal region corresponds 
to the S(IV) subtype associated with ordinary chondrites (OC). The rectangular zone (BA) includes the pyroxene dominated basaltic achondrite assemblages \citep{1993Icar..106..573G}.}}

\end{center}
\end{figure*}

\begin{figure*}[h]
\begin{center}
\includegraphics[height=9cm]{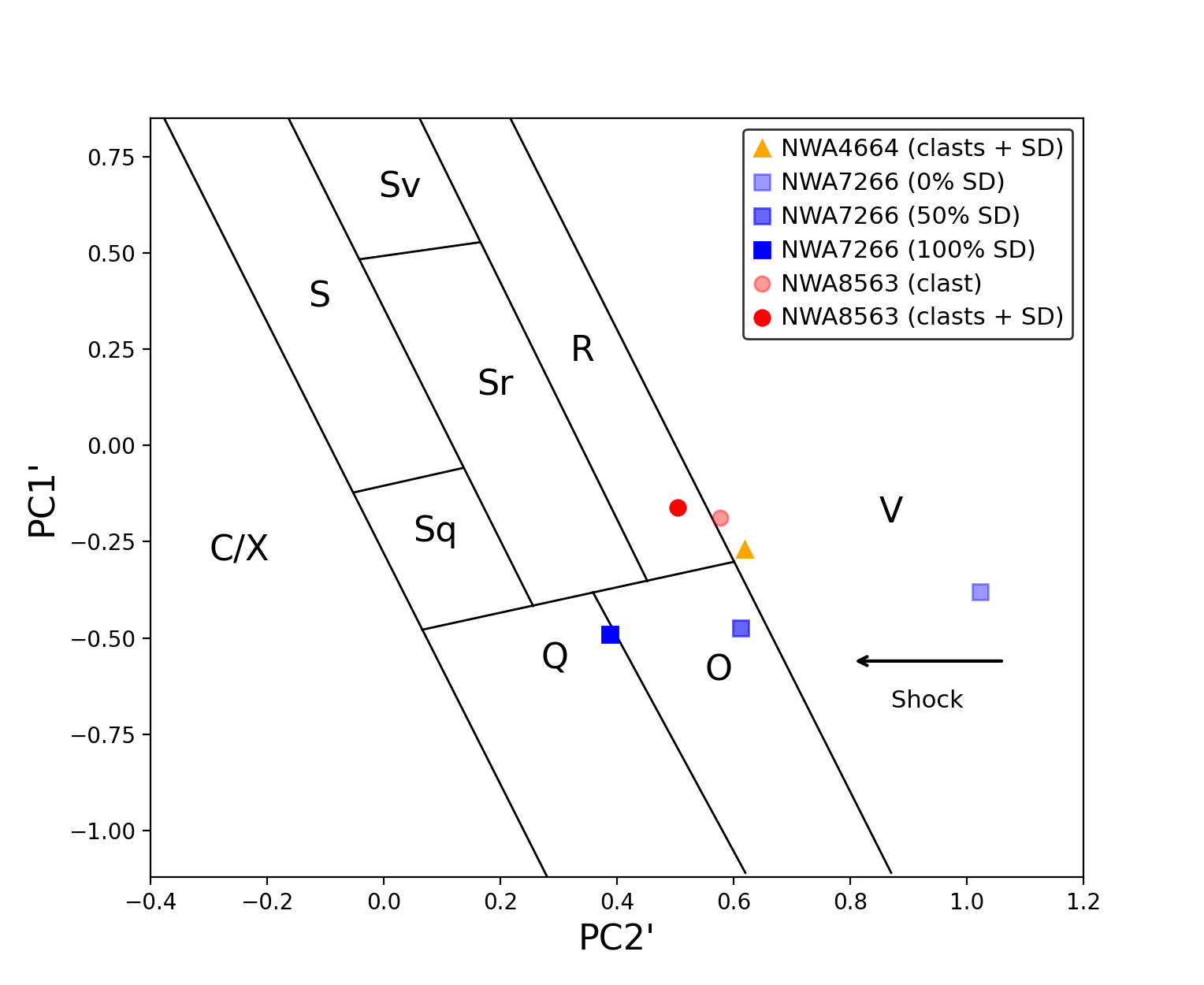}

\caption{\label{f:HED_PC}{\small PC2' vs. PC1' diagram from \cite{2009Icar..202..160D}. PC values calculated for diogenite NWA 4664 and eucrites NWA 7266 and NWA 8563 
corresponding to different fractions of shock darkening (SD) are depicted with different symbols. The increase in SD is represented as an increase in the opacity of the symbols. The arrow
indicates the direction in which shock darkening increases.}}

\end{center}
\end{figure*}

The BAR values of NWA 7266 had the opposite behavior of the ordinary chondrites; i.e, they decreased with increasing shock darkening (Figure \ref{f:BIC_HED}). The initial 
BAR measured for the sample with 0\% shock darkening placed NWA 7266 between the S(V) and S(VI) subtypes. Increasing the fraction of shock-darkened material moved 
NWA 7266 to the S(V) subtype and then close to the S(IV) subtype. The change in taxonomic classification of NWA 7266 was also quite dramatic (Figure \ref{f:HED_PC}). The 
spectrum corresponding to the light-colored lithology was classified as a V-type and after adding 50\% of shock-darkened material was classified as an O-type. The sample 
containing 100\% of shock-darkened material moved further to the left in the PC2' vs. PC1' diagram and was classified as a Q-type. This classification as a Q-type is 
consistent with the location of this sample near the S(IV) subtype (Figure \ref{f:BIC_HED}) which is associated with ordinary chondrites.

\begin{figure} 
\includegraphics[width=9.5cm,angle=0]{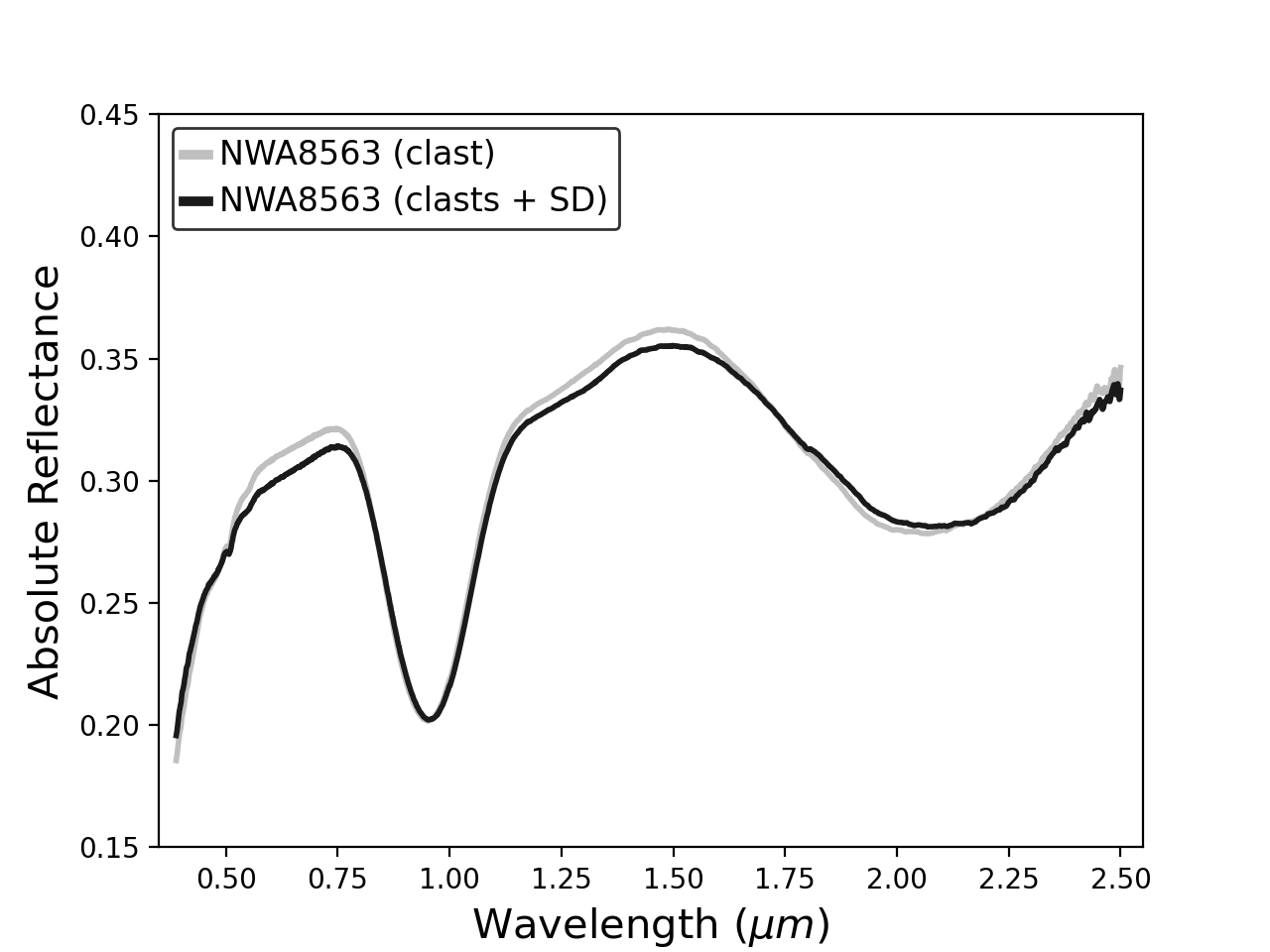}
\hspace{-5mm}
\includegraphics[width=9.5cm,angle=0]{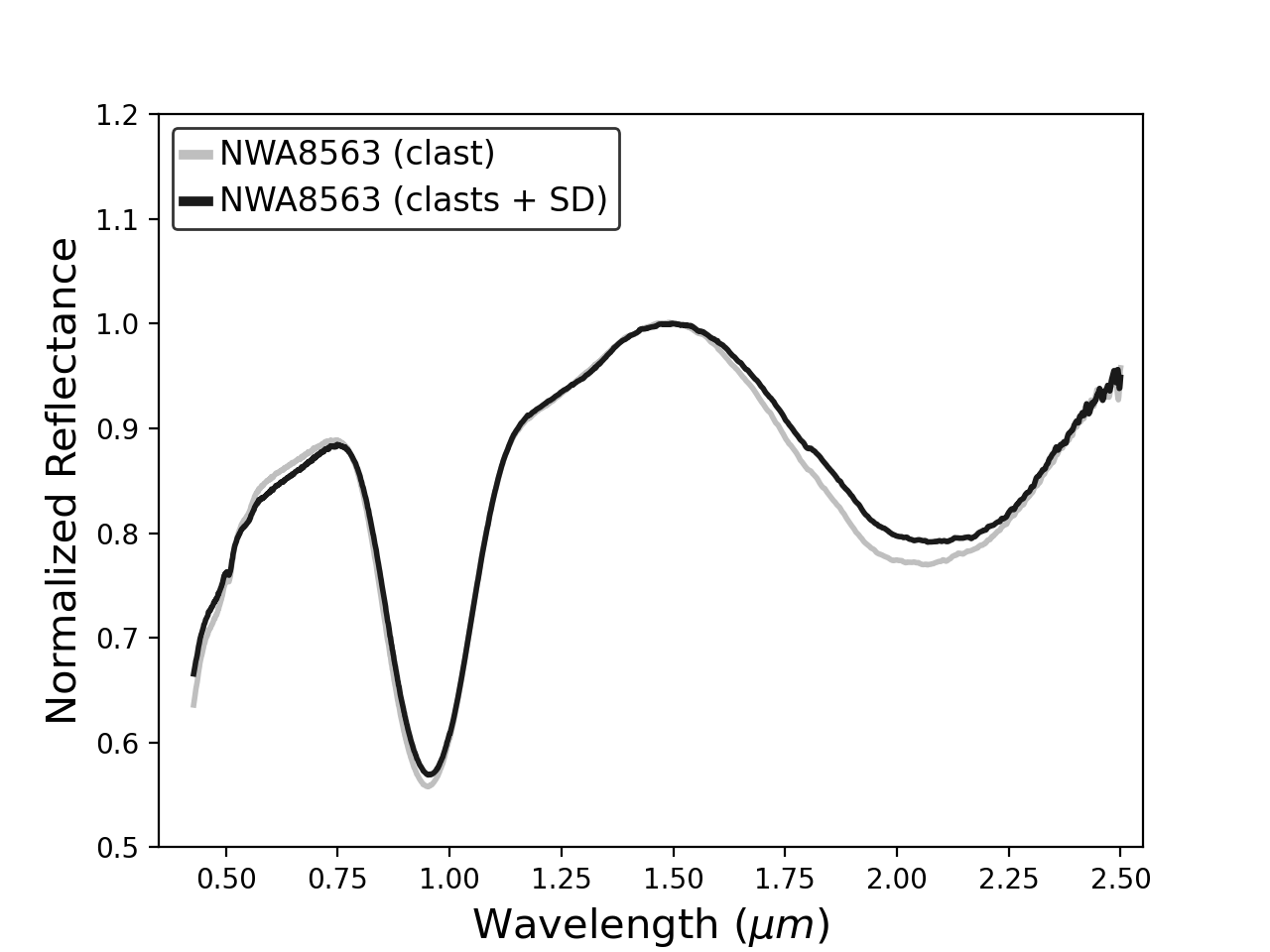}
\caption{Left: Spectra of eucrite NWA 8563 corresponding to the light-brown clast and clasts plus shock darkening (SD) from the dark matrix material. Right: The same spectra normalized to unity at 1.5 $\mu$m.}

\label{f:NWA8563_spec}
\end{figure}

As explained earlier, for eucrite NWA 8563, we were able to prepare one powder made from a light-brown clast and one made from clasts and dark matrix material. 
The spectra of NWA 8563 are shown in Figure \ref{f:NWA8563_spec}. As can be seen, there is little difference between the two samples, with the spectrum 
corresponding to the clasts plus dark matrix material showing a slight decrease in reflectance compared to the spectrum of the light-brown clast. A small increase in 
spectral slope is observed for the spectrum containing dark matrix material, which also shows a small suppression of the absorption bands (Table 1).

The band center values of NWA 8563 shown in Figure \ref{f:BIC_BIIC} suggest that high-Ca clinopyroxene could be present. Similar to NWA 7266, the band centers 
of NWA 8563 shift to longer wavelengths when the shock-darkened material is present, although in this case the shift is small. In the Band I center vs. BAR plot, NWA 
8563 falls in the S(V) subtype (Figure \ref{f:BIC_HED}). We observed a decrease in the BAR in the sample containing dark matrix material; however, it was not enough 
to move NWA 8563 out of the S(V) subtype. The situation was different for the taxonomic classification; the sample made from the light-brown clast was classified as a 
V-type, but the sample containing dark matrix material was classified as R-type even though this spectrum only shows a modest suppression of the absorption bands 
(Figure \ref{f:HED_PC}).

\begin{figure*}[h]
\begin{center}
\includegraphics[height=8cm]{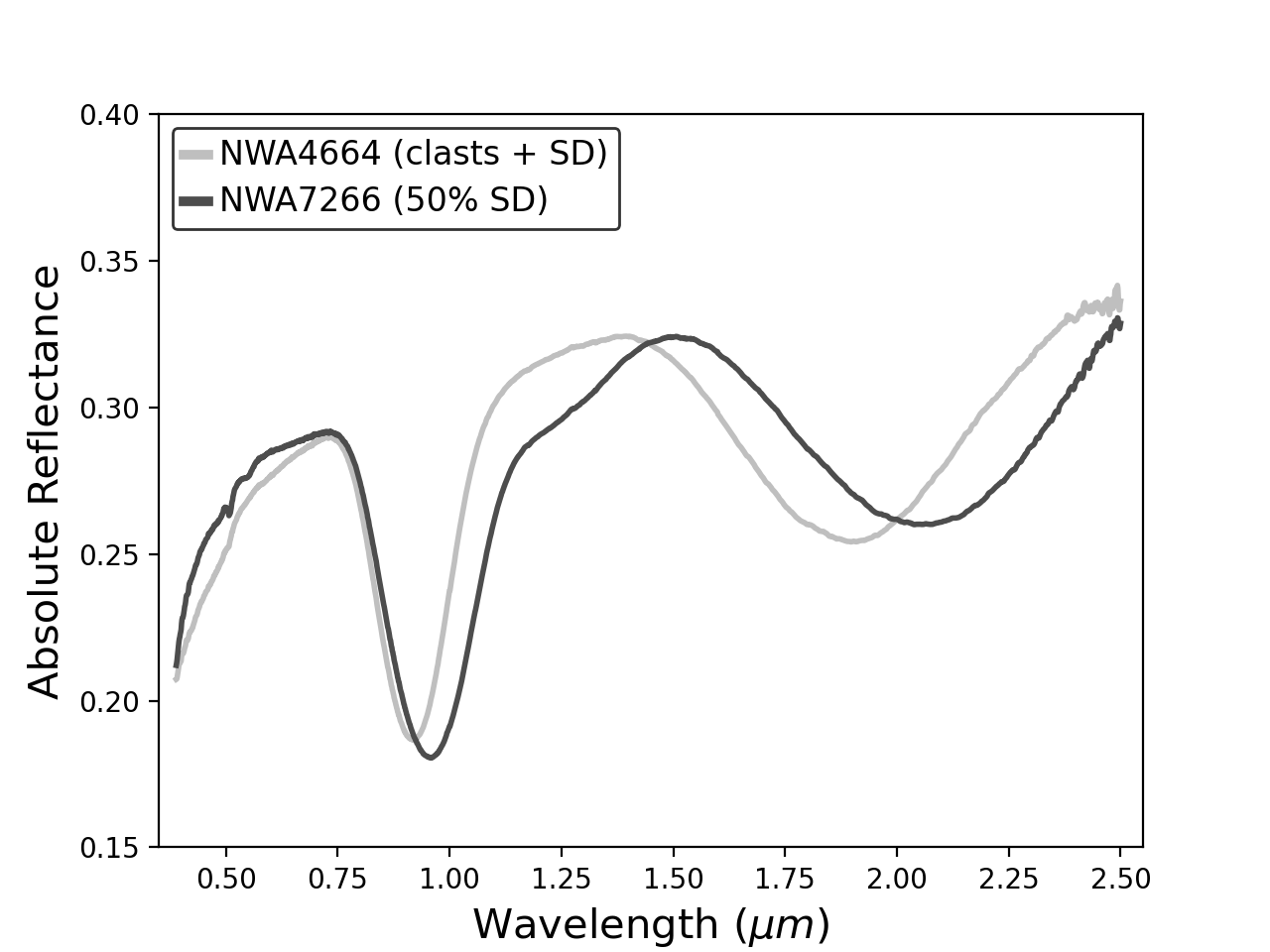}

\caption{\label{f:NWA4664-7266_spec}{\small Spectrum of diogenite NWA 4664 compared to the spectrum of eucrite NWA 7266 corresponding to 50\% shock darkening (SD).}}

\end{center}
\end{figure*}

NWA 4664, the only diogenite, is composed of very small light clasts that are intimately mixed with a dark matrix, which prevented us from 
separating both materials for study. Nevertheless, the spectrum of this sample shows some interesting characteristics. For example, the albedo of NWA 4664 (0.269) is much 
lower than the albedo of the light-colored lithology of NWA 7266 (0.383) and closer to the albedo of NWA 7266 with 50\% shock-darkened material (0.276). Something 
similar happens with the band depths of NWA 4664, which are also closer to the spectrum of NWA 7266 corresponding to 50\% shock darkening. For comparison, both 
spectra are shown in Figure \ref{f:NWA4664-7266_spec}.

The Band I and II centers of NWA 4664 overlap with values found for orthopyroxenes (Figure \ref{f:BIC_BIIC}), which is consistent with the results of \cite{2008M&PS...43..571C}. In the Band I center vs. BAR plot, this meteorite is located between the basaltic achondrite region and 
the S(VII) subtypes (Figure \ref{f:BIC_HED}). The spectrum of NWA 4664 was classified as a V-type, but we noticed that its location in the PC2' vs. PC1' 
diagram is very close to the line that separates V-types from R- and O-types (Figure \ref{f:HED_PC}).

\newpage

\section{Discussion}

Our analysis of ordinary chondrites, eucrites, and diogenites has shown that the most significant changes due to shock darkening occur in the albedo and 
the band depths. For the ordinary chondrites, we found a difference in albedo of $\sim$0.13 between the samples that contain no shock-darkened material and those made of 
100\% shock-darkened material. In the case of the eucrite NWA 7266, the difference in albedo between the two lithologies was $\sim$0.19. To better 
understand the implications of these results, it is useful to look at the typical albedos of S- and V-type asteroids, which have been linked to ordinary chondrites and HEDs, 
respectively. S-types NEOs 
have median albedos of $0.24_{-0.05}^{+0.04}$, while V-types have median albedos of $0.34_{-0.07}^{+0.08}$ \citep{2022AJ....163..165M}. A decrease in albedo of 0.13 for an S-type would make this 
object to fall within the range of albedos calculated for M-types ($0.15\pm0.03$), L-types ($0.12\pm0.04$), B-types ($0.09_{-0.04}^{+0.05}$), and even C-types ($0.04_{-0.01}^{+0.04}$) when the uncertainty of the albedos is considered \citep{2022AJ....163..165M}. Similarly, a decrease in albedo of 0.19 for a 
V-type would make this object resemble other taxonomic types including S-, M-, L-, B- and C-types. Overall, we find that the decrease in albedo produced by 
shock darkening could lead to an overestimation of low albedo objects, such as C-types. These results highlight the limitations of using only albedo 
or colors for taxonomic classification, as is often the case when using photometric data or data obtained by space-based infrared surveys.

It is also worth mentioning that the effects of shock darkening differ from those seen in space weathering. In particular, for objects in the S-complex, space weathering 
produces a distinctive increase in spectral slope and a transition from Q-types to S-types in the PC2' versus PC1' diagram. This transition is represented 
by a space-weathering vector that moves parallel to the line $\alpha$ that separates the S-complex from the C/X-complex \citep[e.g.,][]{2019Icar..324...41B, 2020PSJ.....1...37K}. In contrast, shock darkening 
does not seem to produce a clear trend in the spectral slope and can change the taxonomic type of an asteroid from S- to C/X-complex.

An interesting characteristic of the NEO population is the abundance of objects with an LL chondrite-like composition among S-complex asteroids 
\citep{2008Natur.454..858V, 2013Icar..222..273D, 2014Icar..228..217T, 2019Icar..324...41B, 2024PSJ.....5..131S}. The study done by \cite{2024PSJ.....5..131S} included 56 
NEOs with ordinary chondrite-like compositions and diameters ranging from $\sim$8 to 284 m. In this study, they found that 8\% of the NEOs with ordinary chondrite-like 
compositions were classified as H, 31\% as L, and 61\% as LL chondrites. The high proportion of asteroids with LL-chondrite compositions is inconsistent with the low 
fraction of these meteorites in ordinary chondrite falls where they only represent $\sim$10\%. One possible explanation for this is that the NEOs that have been studied 
are too large to be the immediate parent bodies of the ordinary chondrites that fall on Earth. As a result, they do not share the same compositional trends of these meteorites 
 \citep[e.g.,][]{2008Natur.454..858V, 2019Icar..324...41B, 2024PSJ.....5..131S}. It is also possible that, because of their higher metal content, H and L chondrites are more likely to survive as meteorites than LL chondrites. However, in light of the results obtained in this study, one might wonder whether 
shock darkening could be contributing to the discrepancy. If shock darkening is more prevalent among NEOs with H and L chondrite-like compositions, then our observations 
could be biased against these lower albedo asteroids, making those with an LL chondrite-like composition easier to find. To try to answer this question, we look 
at the statistics associated with shock darkening in ordinary chondrites and NEOs. \cite{1991Metic..26..279B} carried out a survey to determine the abundance of black chondrites and gas-rich ordinary chondrites in the meteorite fall population. A total of 446 black chondrites and 33 gas-rich ordinary chondrites were checked. They 
started by defining a black chondrite meteorite as any ordinary chondrite exhibiting low reflectance $<$0.15 or with major portions of the meteorite containing such low 
reflectance material. Under this deffiniton, they found that black chondrites account for $\sim$13.7\% of ordinary chondrite falls. This number increased to 16.7\% when 
gas-rich chondrites that met the criterion for being black were included. \cite{1991Metic..26..279B} also looked at the distribution of black and gas-rich 
ordinary chondrites between chemical groups and petrographic types. They found that the highest fraction of black chondrites was among H chondrites (16.67\%), followed 
by LL (15.38\%) and L chondrites (11.16\%). In the case of the gas-rich chondrites that could be considered black, they found that the highest fraction corresponds to 
LL (62.50\%) and H chondrites (59.09\%) and the lowest to L chondrites (33.33\%). This distribution indicates that our observations would be more biased against H and 
LL chondrites. For NEOs, obtaining a robust statistic is more difficult due to the limited number of objects that show evidence of shock darkening. \cite{2024PSJ.....5..131S} 
found that $\sim$7\% of the NEOs with ordinary chondrite-like compositions showed possible evidence of shock darkening or high metal content that could explain their weak 
absorption bands. This value is less than half of what was found by \cite{1991Metic..26..279B} (16.7\%) even if we assume that shock darkening is present in all the NEOs 
with subdued absorption bands. The distribution of shock darkening between chemical groups in small NEOs ($<$300 m) is also different. For these objects, the results 
obtained by \cite{2024PSJ.....5..131S} show the highest fraction of possible shock darkening for L chondrites ($\sim$11\%), followed by LL ($\sim$6\%) and H chondrites 
(0\%). 

The differences in abundance and distribution of shock darkening between the meteorites and the NEOs could be due to the limited data available. 
 Furthermore, as explained earlier, because of the size of the NEOs they might not be representative of the immediate parent bodies of the ordinary chondrites. 
What we can say based on the information we have is that, in general, there is a low incidence of black chondrites and NEOs with shock darkening. Although 
there is a higher fraction of shock darkening in L chondrites compared to LL chondrites among NEOs, the difference is too small to explain the abundance 
of objects with an LL chondrite-like composition. Additional data is needed to confirm these results, but considering the observed distribution, we find that shock darkening is unlikely to 
produce a significant bias against the discovery of NEOs with H and L chondrite-like compositions.

\section{Summary} \label{sec:Summ}

We studied the effects of shock darkening on the VNIR spectra of the three subgroups of ordinary chondrites (H, L, LL) and meteorites belonging to the HED clan. When possible, 
intimate mixtures of the light-colored and the shock-darkened lithologies were prepared. For all the samples, we measured the albedo and spectral band parameters and applied 
the Bus-DeMeo taxonomy. The main findings of our work can be summarized as follows:

\begin{itemize}

\item The ordinary chondrites that exhibit a light/dark structure (Chergach, Viñales, and Chelyabinsk) showed a decrease in albedo that 
went from a mean value of 0.205 (no shock-darkened material) to 0.124 (50\% shock-darkened material) and 0.077 (100\% shock-darkened material).

\item No clear trend was observed for the spectral slope of Chergach, Viñales, and Chelyabinsk when the amount of shock-darkened material was increased. However, for all these 
samples, we found a decrease in band depths with increasing shock darkening.

\item We found that the BAR of Chergach, Viñales, and Chelyabinsk increased when the shock-darkened material was added. The increase in BAR for Chergach and Viñales 
resulted in an underestimation of the ol/(ol+px).

\item For Chergach, Viñales, and Chelyabinsk we also found that adding $\gtrsim$50\% of shock-darkened material was enough to change their taxonomic classification from 
S-complex to C/X-complex.

\item Ordinary chondrites Tassédet, NWA 4860, Pampa-C, Ghubara, Tsarev, and Renfrow, which exhibit a uniform dark color, were found to have albedos ranging from 
$\sim$0.07 to 0.14. The high albedo of Chico suggests that there is little or no shock-darkened material present in the sample used in this study. With the exception of Chico and 
Renfrow, all these meteorites were classified in the C- or X-complex. 

\item The albedo of eucrite NWA 7266 decreased with increasing shock darkening, going from 0.383 (0\% shock darkening) to 0.276 (50\% shock darkening) and 0.198 (100\% 
shock darkening). A suppression of the absorption bands with increasing shock darkening was also observed.

\item We found that the BAR values of NWA 7266 had the opposite behavior of the ordinary chondrites, i.e, they decreased with increasing shock darkening. The taxonomic 
classification of this meteorite changed from V-type (0\% shock darkening) to Q-type (100\% shock darkening).

\item The spectrum corresponding to the light clasts plus dark matrix of eucrite NWA 8563 showed little change with respect to the spectrum of the light clast only. However, 
the small decrease in band depths in the spectrum containing dark matrix material was enough to change its taxonomic classification from V-type to R-type.

\item The spectral characteristics of diogenite NWA 4664 in terms of albedo and band depths are similar to those of NWA 7266 containing 50\% of shock-darkened material.

\item The low incidence of black chondrites and NEOs with shock darkening suggest that this process is unlikely to produce a significant bias against the discovery 
of NEOs with H and L chondrite-like compositions. Additional data are needed to confirm these results, but based on the information available, we find that shock darkening is probably not 
responsible for the abundance of LL chondrites among NEOs.

\end{itemize}

\begin{acknowledgments}

This research work was supported by NASA Yearly Opportunities for Research in Planetary Defense Grant 80NSSC22K0514 (PI: V. Reddy). Taxonomic type results presented in this work were determined, in whole or in part, using a Bus-DeMeo Taxonomy Classification Web tool by Stephen M. Slivan, developed at MIT with the support of National Science Foundation Grant 0506716 and NASA Grant NAG5-12355. 
We thank the anonymous reviewers for useful comments that helped improve this paper.

\end{acknowledgments}

\end{document}